\documentclass[apj]{emulateapj}
\usepackage{natbib}

\usepackage{amsmath}
\usepackage{graphicx}
\usepackage{enumitem}
\usepackage{multirow}
\usepackage{color}
\usepackage[dvipsnames]{xcolor}
\usepackage[T1]{fontenc}
\usepackage{ae,aecompl}
\usepackage{amssymb}	
\usepackage{hyperref}
\hypersetup{colorlinks, citecolor=blue, linkcolor=blue}
\def\aj{AJ}
\def\apj{ApJ}
\def\apjl{ApJ}
\def\apjs{ApJS}
\def\aap{A\&A}
\def\mnras{MNRAS}
\def\nat{Nature}
\def\pasp{PASP}%
\def\araa{ARA\&A}

\def\pasj{PASJ}
\begin{document}

\title{Stellar Populations of Nine Passive Spiral Galaxies from the CALIFA survey: are they progenitors of S0{\scriptsize s}?}

\author{Mina Pak\altaffilmark{1,2}, 
Joon Hyeop Lee\altaffilmark{1,2},
Hyunjin Jeong\altaffilmark{1}, 
Suk Kim\altaffilmark{1,3}, 
Rory Smith\altaffilmark{1},
Hye-Ran Lee\altaffilmark{1,2}
}

\affil{\altaffilmark{1}Korea Astronomy and Space Science Institute (KASI), 
   776 Daedukdae-ro, Yuseong-gu, Daejeon  34055, Republic of Korea}
\affil{\altaffilmark{2}University of Science and Technology, Korea (UST), 
   217 Gajeong-ro Yuseong-gu, Daejeon 34113, Republic of Korea}
\affil{\altaffilmark{3}Center for Galaxy Evolution Research, Yonsei University,
   Seoul, 03722, Republic of Korea}
   
\email{minapak@kasi.re.kr}

\begin{abstract}
We investigate the stellar population properties of passive spiral galaxies in the CALIFA survey. Nine spiral galaxies that have NUV$-$r $> 5$ and no/weak nebular emission lines in their spectra are selected as passive spirals. Our passive spirals lie in the redshift range of $0.001 <$ z $< 0.021$ and have stellar mass range of $10.2 <$ log(M$_{\star}/$M$_{\odot}) < 10.8$. They clearly lie in the domain of early-type galaxies in the WISE IR color-color diagram. We analyze the stellar populations out to two effective radius, using the best-fitting model to the measured absorption line-strength indices in the Lick/IDS system. We find that stellar populations of the passive spirals span a wide range, even in their centers, and hardly show any common trend amongst themselves either. We compare the passive spirals with S0s selected in the same mass range. S0s cover a wide range in age, metallicity, and $[\alpha/$Fe], and stellar populations of the passive spirals are encompassed in the spread of the S0 properties. However, the distribution of passive spirals are skewed toward higher values of metallicity, lower [$\alpha$/Fe], and younger ages at all radii. These results show that passive spirals are possibly related to S0s in their stellar populations. We infer that the diversity in the stellar populations of S0s may result from different evolutionary pathways of S0 formation, and passive spirals may be one of the possible channels.

\end{abstract}

\section{Introduction}
Ever since the discovery of significant number of passive spiral galaxies without any apparent sign of ongoing star formation, the formation and evolution of these galaxies have received considerable attention observationally and theoretically (\citealt{Cou98}; \citealt{Dre99}; \citealt{Pog99}; \citealt{Bek02}; \citealt{Got03}; \citealt{Yam04}; \citealt{Mor06}; \citealt{Cor09}; \citealt{Hug09}; \citealt{Lee08}; \citealt{Mah09}; \citealt{Wol09}; \citealt{Mas10}; \citealt{Fra16}; \citealt{Fra18}). The first unusual spiral galaxies were identified by \citet{van76} in the Virgo cluster, so-called `anemic' spirals, and later, passive spirals were found in all environments and at low to high redshifts. Since early samples of red or passive spirals often show the evidence of mild star formation (SF) such as weak nebular emissions, ultraviolet (UV) light from young stellar populations, and infrared (IR) excess from warm dust by obscured SF, some of them may be in progress of star formation quenching rather than completely quiescent. On the other hand, some galaxies show little or no star formation activity despite their obvious spiral structure -- these are the genuine passive spirals. These unique galaxies are clear outliers from the well known color-morphology relations (\citealt{Hub38}; \citealt{Fio99}; \citealt{Ber00}; \citealt{Mig09}). 

The existence of these galaxies leads to the following questions: (1) What is/are the origin(s) of passive spirals? (2) What physical processes have halted the star formation in passive spirals without destroying their spiral structures? (3) Are passive spirals eventually transformed into lenticular galaxies? 

Recent studies suggest that the formation of passive spirals in the intermediate to high density environments is closely related with cluster environmental effects (\citealt{Got03}; \citealt{Mas10}; \citealt{Fra18}) such as thermal evaporation \citep{Cow77}, ram pressure stripping \citep{Gun72}, harassment \citep{Moo99}, starvation or strangulation (\citealt{Lar80}; \citealt{Bek02}), galaxy-galaxy interactions in high density regions including major \citep{Too72} and minor \citep{Wal96} mergers and tidal interactions. Moreover, a very recent study of \citet{Fra18} found that all low mass (M$_\star$ $< 1 \times 10^{10} M_{\odot}$) passive spirals in their sample are members of the Virgo cluster, and thus cluster-related environmental phenomena are most likely responsible for the quenching of those galaxies.

The existence of passive spirals is the strong evidence for the  morphological transformation of spirals into S0s in rich clusters. \citet{Bek02} showed how cluster environmental quenching processes can transform spirals into S0s, passing through an intermediate passive spiral phase, using numerical simulations. After the gas is stripped, the spiral arm structures fade over several Gyrs. However, they claim that passive spirals are found anywhere from isolation to the centers of clusters and hence that no single mechanism can completely explain their origins. 

In low density environments, secular evolution as a result of bars is a viable mechanism for quenching spirals (\citealt{Kor79}; \citealt{Kor04}; \citealt{Ath13}), which may convert them into S0s. Bars are commonly found structures in disk galaxies of the local universe. Recent work on the bar fraction of nearby disk galaxies suggests that the bar fraction is up to $\sim 50\%$ in optical studies (\citealt{Mar07}; \citealt{Ree07}; \citealt{Bar08}). The fraction rises to about 70\% in near-infrared studies (\citealt{Kna00}; \citealt{Men07}). 

Bars play a significant role in the evolution of galaxy structure and morphology by transferring angular momentum and energy, and redistributing mass. They act on both baryonic and dark matter components of a galaxy (\citealt{Wei85}; \citealt{Deb98}, \citealt{Deb00}; \citealt{Ath03}). Bars funnel material towards the galaxy center, helping to grow bulge-like structures (\citealt{Kor04}), nuclear bars (\citealt{Erw04}) and feeding the central black hole (\citealt{Shl89}; \citealt{Shl90}; \citealt{Shl00}; \citealt{Jog06}). In particular, the role of driving gas inwards can trigger star formation in the center (e.g. \citealt{Haw86}; \citealt{Jog05}; \citealt{She05}), and lead to the more rapid consumption of the gas. \citet{Mas12} showed that a higher bar fraction in disk galaxies is associated with a lower HI gas content, and there are hints that at a fixed HI content barred galaxies are optically redder than unbarred galaxies. Since high-mass passive spirals (M$_\star$ $> 1 \times 10^{10} M_{\odot}$) are found not only in high density regions but in very low density regions (\citealt{Bam09}; \citealt{Mas10}; \citealt{Fra18}), it is suggested that internal processes such as bars are more important mechanisms to quench the star formation in high mass passive spirals. 

While all these devoted studies have drawn attention to passive spirals, the origins and evolutionary pathways of passive spirals are still under debate. To confirm the origin of passive spirals, strong constraints can be provided by investigating the internal distribution of their stellar populations, which are poorly understood so far. That is, an important clue to their origin can be found in whether the quenching is inside-out or outside-in, as well as how old these systems are and how recently their star formation is quenched. Comparing the resolved stellar populations of passive spirals with S0s, their probable descendants, will be helpful to constrain the specific transformation mechanisms that are in action. In addition, the environmental dependence of passive spirals is also unclear. Thus, investigating the environments of individual galaxies may also provide constraints on the quenching mechanisms.

The purpose of this paper is to better understand the evolution of passive spiral galaxies by investigating their spatially resolved stellar populations, using the Calar Alto Legacy Integral Field Area (CALIFA; \citealt{San12}; \citealt{Hus13}) survey, which has been observing a wide range of galaxy morphologies with a wide field-of-view (FoV). In the last decade, the advent of integral field units (IFU) has allowed us to study the details of integrated and spatially resolved spectroscopic properties of galaxies, based on two-dimensional (2D) maps of kinematics and stellar populations. This is the first study of spatially resolved stellar populations for passive spirals. In this study, we ultimately aim to find evolutionary links between passive spirals and S0s.

This paper is organized as follows. Section 2 describes the galaxy samples adopted from the CALIFA survey. The analysis of data including line measurement is in Section 3. We present our results of stellar population properties for the passive spirals in Section 4. Finally, our discussion and conclusions on the evolution of passive spirals are given in Section 5. 

\section{DATA and Sample Selection}
The CALIFA survey (\citealt{San12}; \citealt{Hus13}) was carried out using the 3.5m telescope of the Calar Alto observatory using the PMAS/PPAK spectrograph (\citealt{Rot05}; \citealt{Kel06}). The field of view of PPAK is $74\arcsec$ $\times$ 64$\arcsec$, which is filled with 382 fibers of 2.7$\arcsec$ diameter each \citep{Kel06}. The galaxies were observed with two spectroscopic setups, using the gratings V500 with a nominal resolution ($\lambda$/$\Delta \lambda$) of 850 at 5000 {\AA} (FWHM $\sim$ 6 {\AA}) and a wavelength range from 3745 to 7500 {\AA}, and V1200 with a better spectral resolution of 1650 at 4500 {\AA} (FWHM $\sim$ 2.7 {\AA}) and ranging from 3650 to 4840 {\AA}. More detailed information about the CALIFA sample, and the observational strategy, are available in the papers of CALIFA team (\citealt{Wal14}; \citealt{San12}; \citealt{Hus13}; \citealt{Gar15}; and \citealt{San16}). We here analyzed galaxies using only the V500 data cube for consistency. 

The classification is based on the visual inspection by the authors using the SDSS g$-$, r$-$, and i$-$band composite color images and the spectra. We select spirals with red optical color and without nebular emissions. Among the visually selected red spirals, we limit the NUV$-$r colors of our sample to be larger than $5$ using the Galaxy Evolution Explorer (GALEX) in order to exclude actively star forming galaxies. In addition, edge-on galaxies, shell galaxies, and galaxies with ongoing mergers or strong tidal interactions with neighbors are excluded from our sample. 

\begin{figure}
\includegraphics[width=8.5cm]{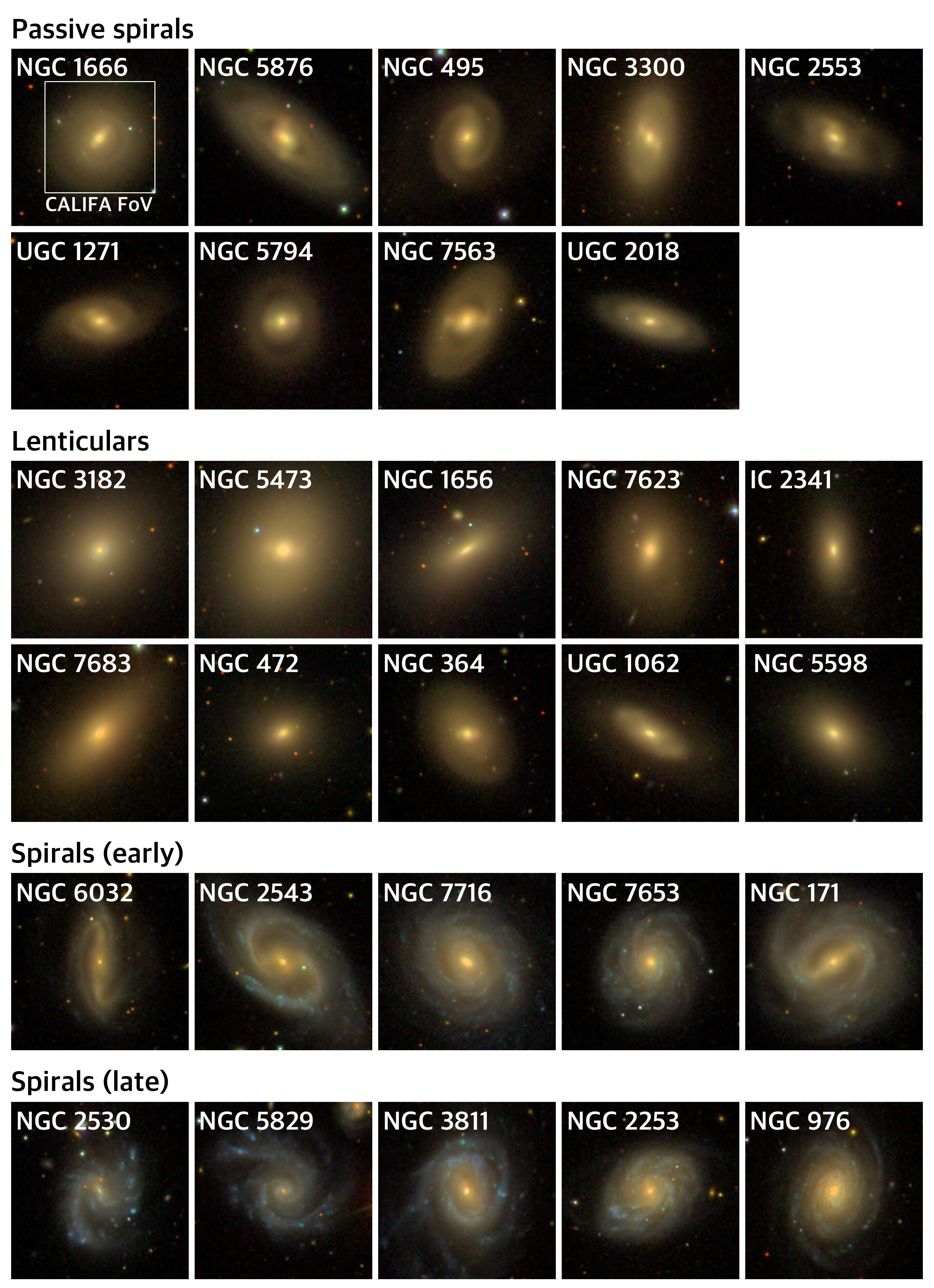}
\caption{The SDSS g$-$, r$-$, and i$-$bands composite images of our 9 passive spirals and the example of S0 and spiral comparisons from the CALIFA archive. The galaxies are ordered from low to high stellar mass according to their NSA catalog stellar mass, for each morphology separately. All images are 120$\arcsec$ $\times$ 120$\arcsec$ in size. North is at the top and east is to the left. The white box in NGC1666 is CALIFA field of view.
\label{F1}}
\end{figure}

This process yields a final sample of $9$ passive spiral galaxies. The SDSS images of these galaxies are presented in Figure \ref{F1}. Six of them show well-developed spiral arms. The others have less prominent spiral arms but have subtle structures with discontinuous stellar distributions in their disks. We present about half of S0s in our sample to show that they are clearly different from passive spirals. S0s would have central star forming ring, inner bar and X-shaped structures but no spiral arms.

\begin{figure}
\includegraphics[width=8.5cm]{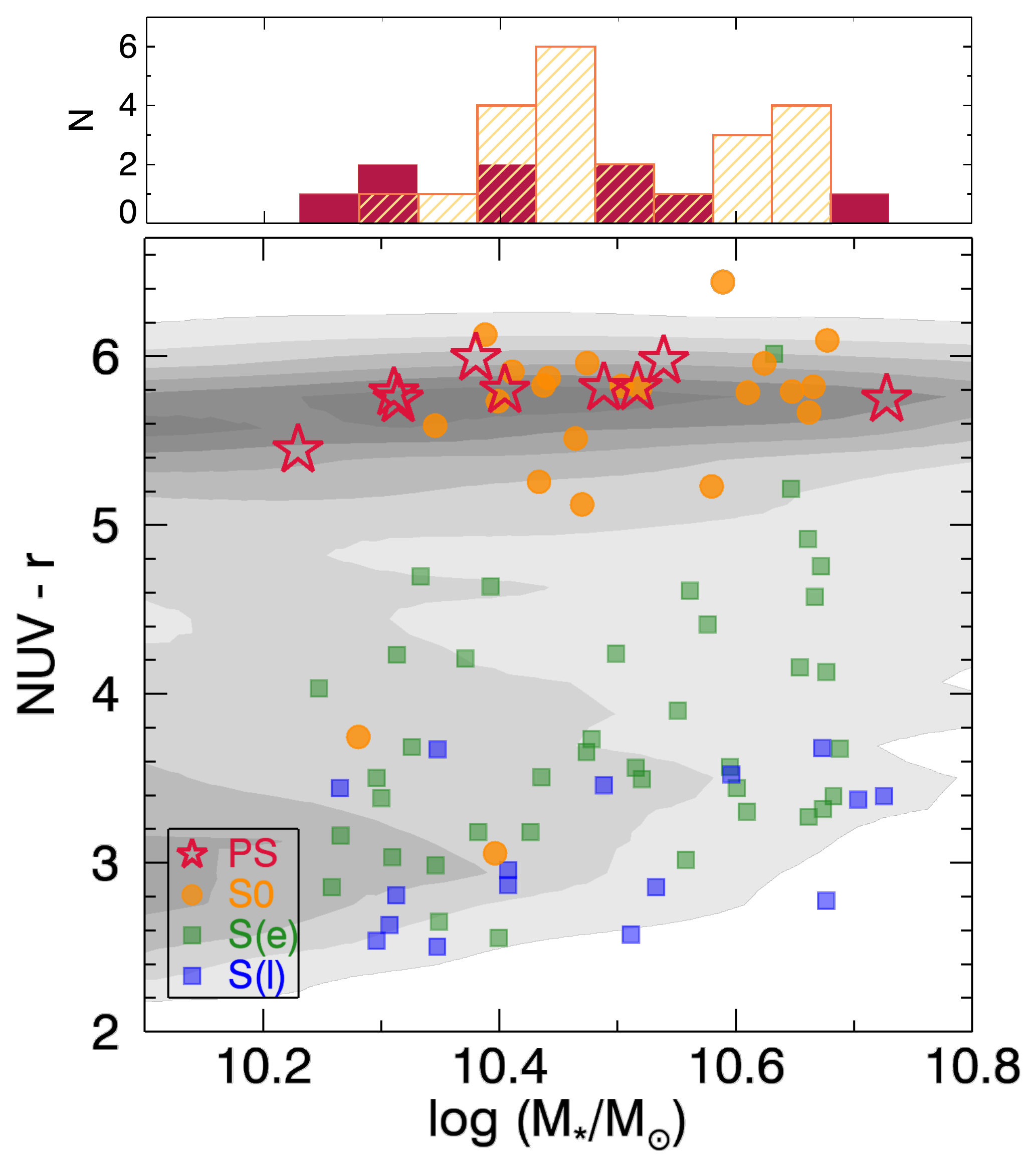}
\caption{NUV$-$r vs. log (M$_{\star}/$M$_{\odot}$) for passive spirals and S0s. Red stars (and red in histogram) are 9 passive spirals and orange circles (and orange hatched in histogram) are 21 S0s. The green and blue squares indicate early- and late-type star-forming spiral galaxies. The distribution of galaxies from the NSA catalog is overlaid as contours.
\label{F2}}
\end{figure}

\begin{figure}
\includegraphics[width=8.5cm]{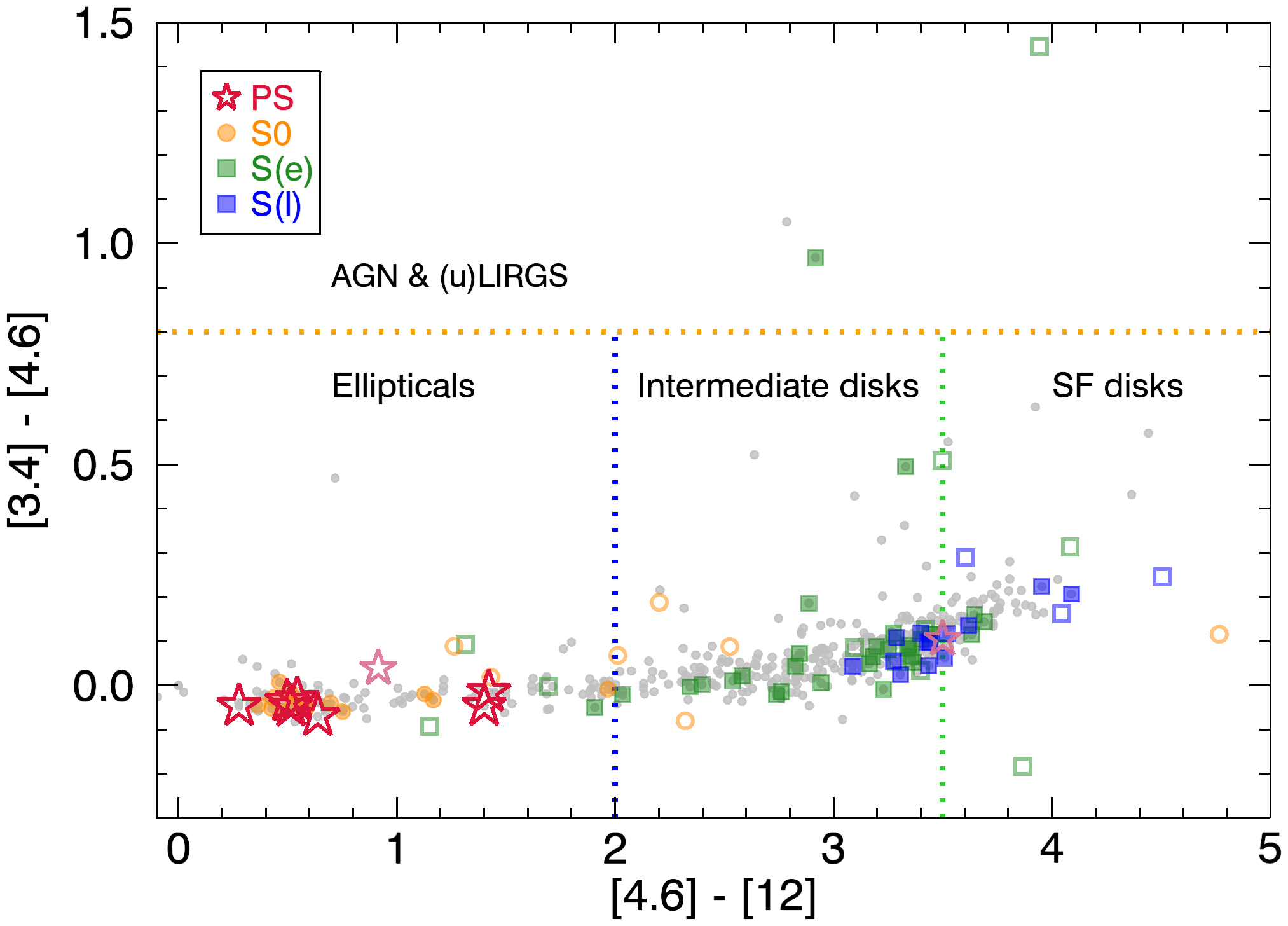}
\caption{WISE color-color diagram for all CALIFA galaxies. The spheroidal (early-type), intermediate disk (disk), and star forming (SF) disk (late-type disk) galaxies are divided by dotted lines based on the schematic diagram from \citet{Jar17}. Symbols are the same as those in Figure \ref{F2}. We use the magnitudes in the WISE catalog extracted using elliptical aperture photometry. When the elliptical-aperture magnitude is not available, we use the instrumental profile-fit photometry magnitude (light red stars for passive spirals,  orange open circles for S0s, green and blue open squares for early- and late-type spirals, respectively).
\label{F3}}
\end{figure}

Passive spirals have stellar masses of $10.23 \lesssim$ log (M$_{\star}/$M$_{\odot}) \lesssim 10.8$, based on stellar masses from the NASA-Sloan Atlas (NSA, \citealt{Bla11}) catalog (Figure \ref{F2}). All passive spirals but one (NGC 3300) have $4.6 \micron -12 \micron < 2$ in the Wide-field Infrared Survey Explorer (WISE) IR color-color diagram (red stars in Figure \ref{F3}), which is the region where typical early-type galaxies are located (see Fig. 11 (b) of \citealt{Jar17}). Table \ref{Tbl1} summarizes the basic information of the 9 passive spirals.

From the CALIFA sample we select lenticular and star-forming spiral galaxies in the same stellar mass range as our passive spiral sample in order to more  fairly compare their stellar populations. For our star forming spiral sample, we adopt  the subtype of spirals from the NASA/IPAC Extragalactic Database (NED). Like the passive spiral sample, we excluded ongoing mergers or strong tidal interactions with neighbors. This results in a sample of $21$ S0s and $57$ spirals as a comparison sample. About half of S0s show embedded structures such as a nuclear star forming ring, inner bars/spirals, rings and X-shapes in disk, features that are typically observed in S0s (\citealt{But13}).

\begin{table*}   
\caption{Basic information of 9 passive spirals}
 \label{Tbl1}
\begin{tabular}{lccccccccrc}
\hline
Name     & R.A.      & DEC.        & Redshift & log (M$_{\star}/$M$_\odot$)  & M$_r$   & NUV$-$r   & $[4.6]-[12]$ &  $[3.4]-[4.6]$ & R$_e$   & n   \\
         & (deg.)    & (deg.)      &          &                              & (mag)   &           &              &                & ($\arcsec$) &   \\
 (1)     & (2)       & (3)         & (4)      & (5)                          & (6)     & (7)       & (8)          &  (9)           & (10)        &  (11)    \\

\hline
NGC 1666 &  072.1368 &  -06.570053 &  0.001   &  10.23  &  -20.16  &  5.44  &  $1.42 \pm 0.049$ &  $-0.02 \pm 0.009$  &   6.73  & 5.76  \\
NGC 5876 &  227.3816 &   54.506515 &  0.011   &  10.31  &  -20.36  &  5.78  &  $1.40 \pm 0.014$ &  $-0.05 \pm 0.007$  &  10.51  & 5.14  \\
NGC 5794 &  223.9734 &   49.726063 &  0.014   &  10.31  &  -20.37  &  5.74  &  $0.50 \pm 0.044$ &  $-0.04 \pm 0.008$  &   5.91  & 4.69  \\
NGC 495  &  020.7331 &   33.471387 &  0.014   &  10.38  &  -20.54  &  5.99  &  $0.51 \pm 0.069$ &  $-0.05 \pm 0.009$  &   9.79  & 4.52  \\
NGC 3300 &  159.1601 &   14.171126 &  0.010   &  10.40  &  -20.59  &  5.80  &  -                &   -                 &  12.55  & 2.93  \\
NGC 2553 &  124.3958 &   20.903057 &  0.016   &  10.45  &  -20.78  &  5.81  &  $0.28 \pm 0.109$ &  $-0.05 \pm 0.011$  &   8.30  & 4.51  \\
NGC 7563 &  348.9830 &   13.196131 &  0.014   &  10.55  &  -20.87  &  5.81  &  $0.64 \pm 0.039$ &  $-0.07 \pm 0.009$  &   8.77  & 3.84  \\
UGC 1271 &  027.2502 &   13.211104 &  0.017   &  10.54  &  -20.93  &  5.97  &  $0.54 \pm 0.050$ &  $-0.03 \pm 0.009$  &   8.20  & 4.20  \\
UGC 2018 &  038.1673 &   00.260080 &  0.021   &  10.73  &  -21.47  &  5.74  &  -                &   -                 &   8.70  & 5.18  \\
\hline
\end{tabular}
\\
\begin{flushleft}
Notes. Column 1 shows the name of the passive spiral, and Column 2 and 3 show the coordinates. Column 4 gives redshift, while Column 5 and 6 give the stellar mass and absolute magnitude in r$-$band, respectively. Column 7, 8 and 9 denote NUV$-$r and IR colors, respectively. Column 10 and 11 show the effective radius and S\'ersic index, respectively. Column 1 to 7, and Column 10 and 11 are from the NSA catalog. The NUV and r absolute magnitudes in the NSA are given for $h = 1$, where H$_0 = 100h$ km s$^{-1}$ Mpc$^{-1}$. IR colors in Column 8 and 9 are from the WISE catalog.
\end{flushleft}
\end{table*}


\section{Analysis} 
We spatially bin the CALIFA data by means of the centroidal Voronoi tessellation algorithm of \citet{Cap03} using the PINGSoft \citep{Ros12} software. The minimum S/N (per bin) for each galaxy is set to be between $15$ and $20$. To derive the stellar velocity ($v_{star}$) and velocity dispersion ($\sigma_{\star}$), emission lines are masked and the absorption line spectrum is fitted with the penalised pixel-fitting approach \citep[pPXF;][]{Cap03} using the stellar population model templates from the MILES stellar library \citep{Vaz10}. The process includes spatial masking of sky lines and bad pixels. The best fitting template combination is determined by $\chi^2$ minimization in pixel space. The best values of radial velocities and absorption line broadening ($\sigma$) for the stellar component are derived in each binned spectrum. 

\subsection{Lick Indices}
The spectral range of the CALIFA data covers a number of absorption lines that can be used as tools to probe the stellar populations of a galaxy. We quantify the absorption features using line-strength indices such as H$\beta$, Mg$b$, Fe5270 and Fe5335, taken from the Lick/IDS system, using the code \textit{lick\_ew.pro} of \citet{Gra08}, part of their {\scriptsize EZ\_AGES} package, which is widely used to measure stellar populations of galaxies and star clusters from Lick indices.

Before calculating line-strength indices, we broadened the CALIFA spectra so that their spectral resolution after velocity dispersion correction in each bin agrees with the spectral resolution of each index in the Lick system (\citealt{Wor97}; \citealt{Tra98}) as implemented in the \textit{lick\_ew.pro} routine. We correct the intrinsic H$\beta$ emission to obtain the `genuine' H$\beta$ absorption line strength, because the observed H$\beta$ line is the combined result of intrinsic emission and intrinsic absorption. We regard the difference in H$\beta$ absorption line strength between the observation and best-fit template as the intrinsic H$\beta$ emission, and the genuine H$\beta$ absorption line-strength is recalculated by adding up this difference. The correction is applied only when the amplitude-to-noise (A/N) of H$\beta$ in each bin is greater than $3$. The estimated intrinsic emission values are similar to typical measurement errors at each radius. For passive spirals, the typical measurement error and intrinsic H$\beta$ emission are $0.17$ and $0.15$ {\AA} at the center and $0.25$ and $0.26$ {\AA} at 1 R$_e$. 

\subsection{Lick Index Grid Method} 

To derive luminosity-weighted simple stellar population (SSP)-equivalent parameters, we use the Lick index grid method, which derives the age, metallicity and abundance ratio by comparing Lick indices to SSP model grids, with iterations between different index-index planes \citep{Puz05}. Here, we adopt SSP models given by \citet{TMB} (hereafter, TMB model).

We use the indices plane of [MgFe]$^\prime$\footnote{[MgFe]$^\prime$ $=$ $\sqrt{\textrm{Mg}b \times (0.72 \times \textrm {Fe}5270 + 0.28 \times \textrm{Fe}5335)}$} versus H$\beta$ to estimate the metallicity and age. The [MgFe]$^\prime$ is insensitive to [$\alpha$/Fe], so it is a good tracer of metallicity. H$\beta$ is an age indicator, which is the least sensitive to [$\alpha$/Fe] among the Balmer lines \citep{Tho03}. We also use the Mg$b$-<Fe>\footnote{<Fe> $=$ (Fe5270 + Fe5335)$/2$} plane to estimate [$\alpha$/Fe]. The Mg$b$ is sensitive to [$\alpha$/Fe], and the <Fe> is a metallicity indicator.

To obtain fine accuracy in the estimation, we interpolate the  TMB models to grids with $\sim 300,000$ individual models, spanning [Z/H] $= -0.33$ to $0.67$ (in the interval with $30$ steps), age $= 1$ to $15$ Gyr ($140$ steps) and [$\alpha$/Fe] $= -0.3$ to $0.5$ ($30$ steps). The majority of our measurements for the galaxies in this study are covered by these model parameter ranges. For seven points that lie outside the model grid space, we use the best fitting values at the boundary of the interpolated models.

\begin{figure*}
\centering
\includegraphics[width=18cm]{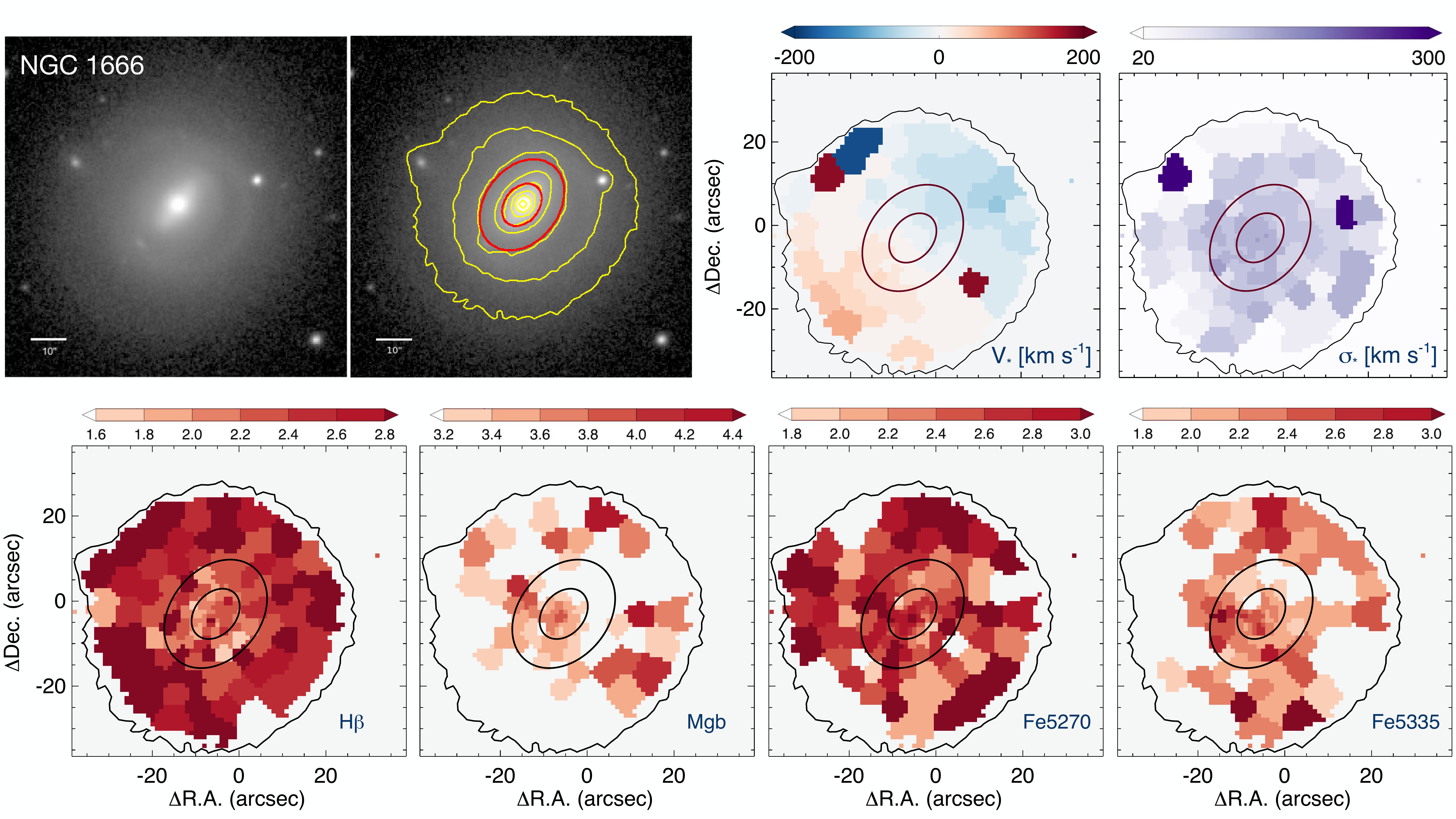}
\caption{The SDSS optical images and maps of the derived parameters of the $9$ passive spirals in the CALIFA sample, ordered by increasing stellar mass from the NSA catalog. In the upper row, from left to right we show the SDSS r-band image, the SDSS r-band image with surface brightness contours (in yellow) and the one and two effective radius (ellipses), the stellar velocity map, and the velocity dispersion map. The velocity is normalized relative to the velocity of the brightest pixel. In the bottom row, we present the H$\beta$, Mg$b$, Fe5270 and Fe5335 maps. We mark the outermost of the surface brightness contour and the one and two effective radius (ellipses) on the parameter maps. In order to easily compare between galaxies, the range of the color bars is fixed for each parameter.
\label{F41}}
\end{figure*}

\addtocounter{figure}{-1}

\begin{figure*}
\centering
\includegraphics[width=18cm]{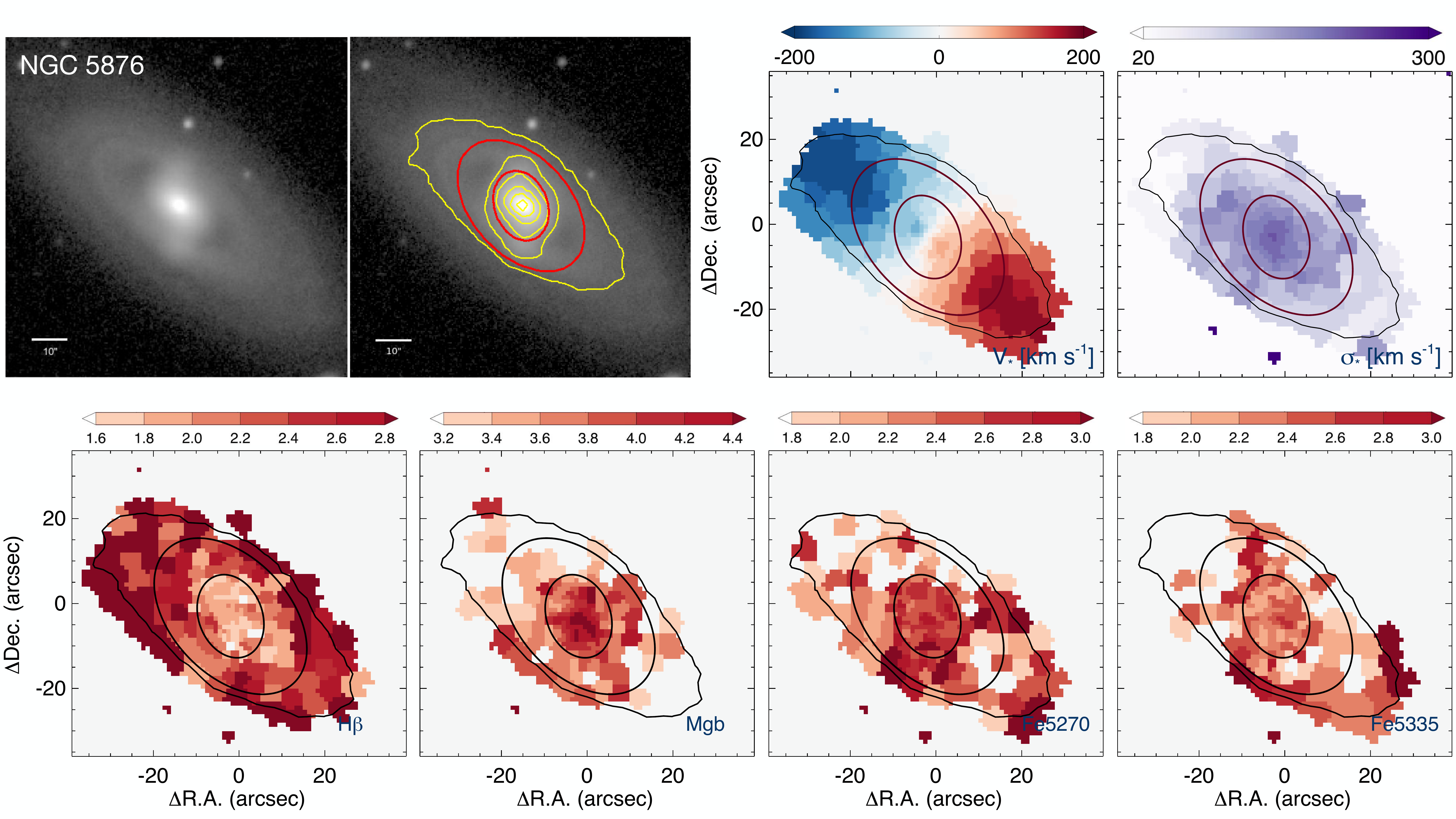}
\caption{ {\it continued}
\label{F42}}
\end{figure*}

\addtocounter{figure}{-1}

\begin{figure*}
\centering
\includegraphics[width=18cm]{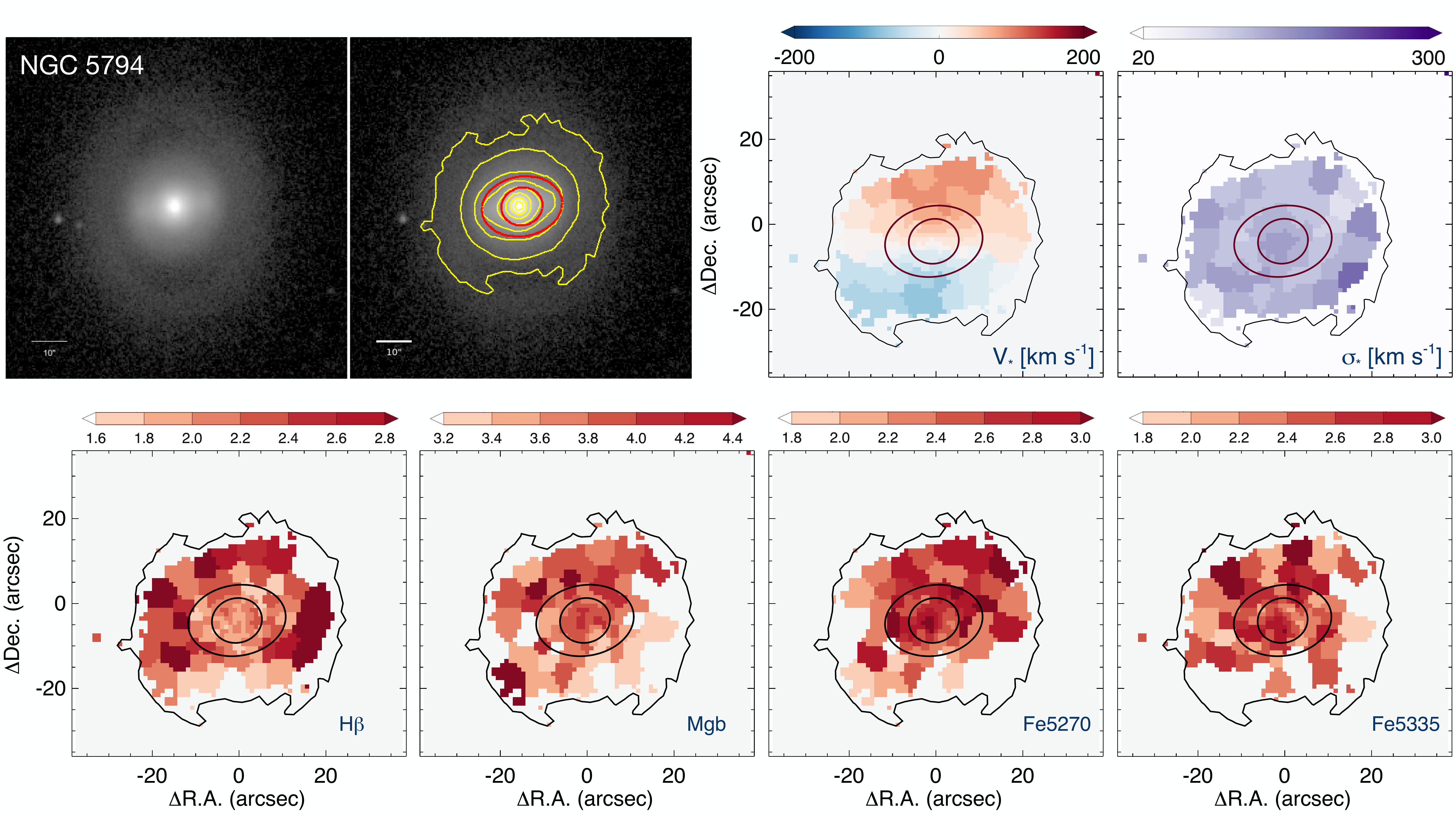}
\caption{ {\it continued}
\label{F43}}
\end{figure*}

\addtocounter{figure}{-1}

\begin{figure*}
\centering
\includegraphics[width=18cm]{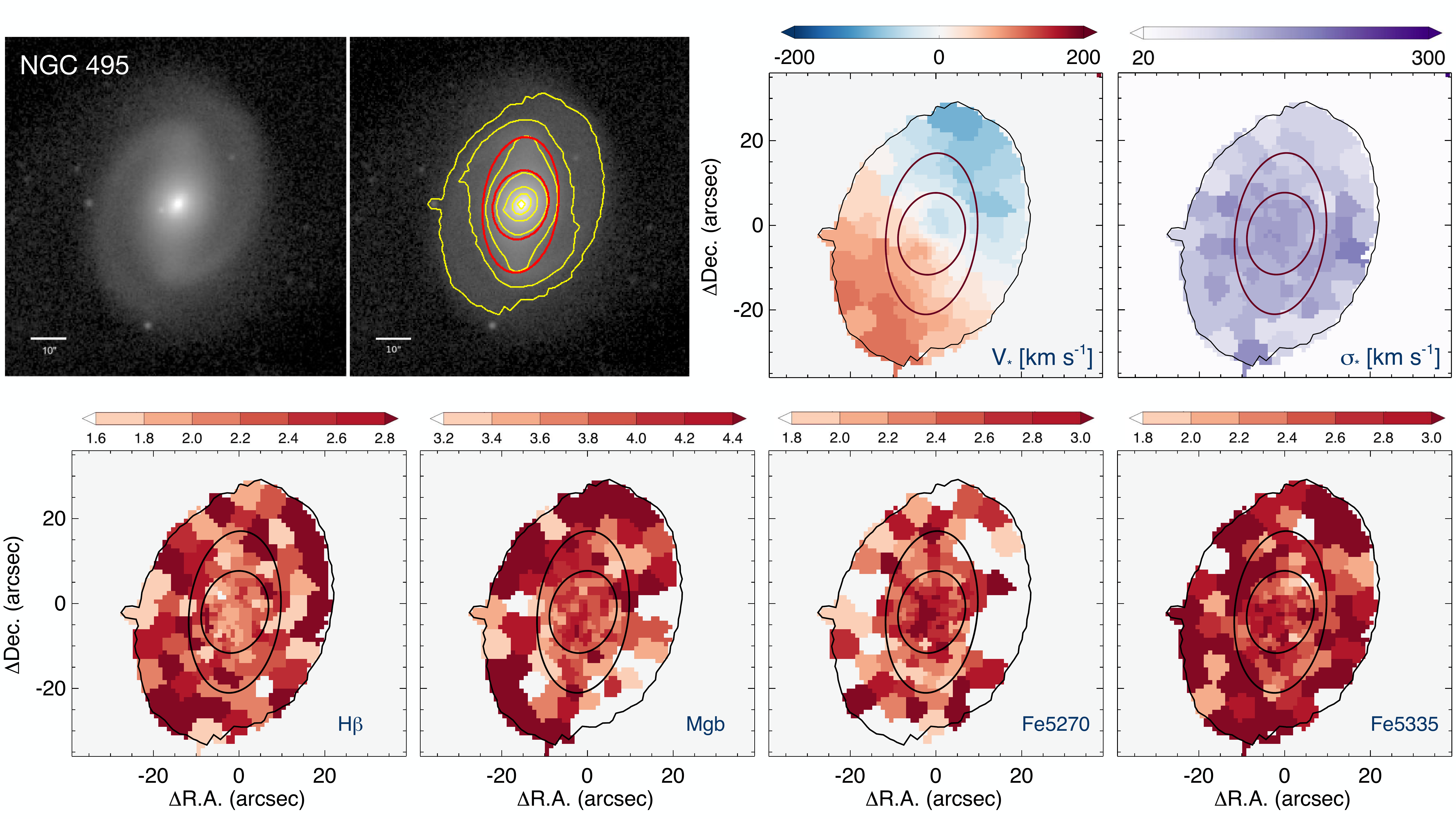}
\caption{ {\it continued}
\label{F44}}
\end{figure*}

\addtocounter{figure}{-1}

\begin{figure*}
\centering
\includegraphics[width=18cm]{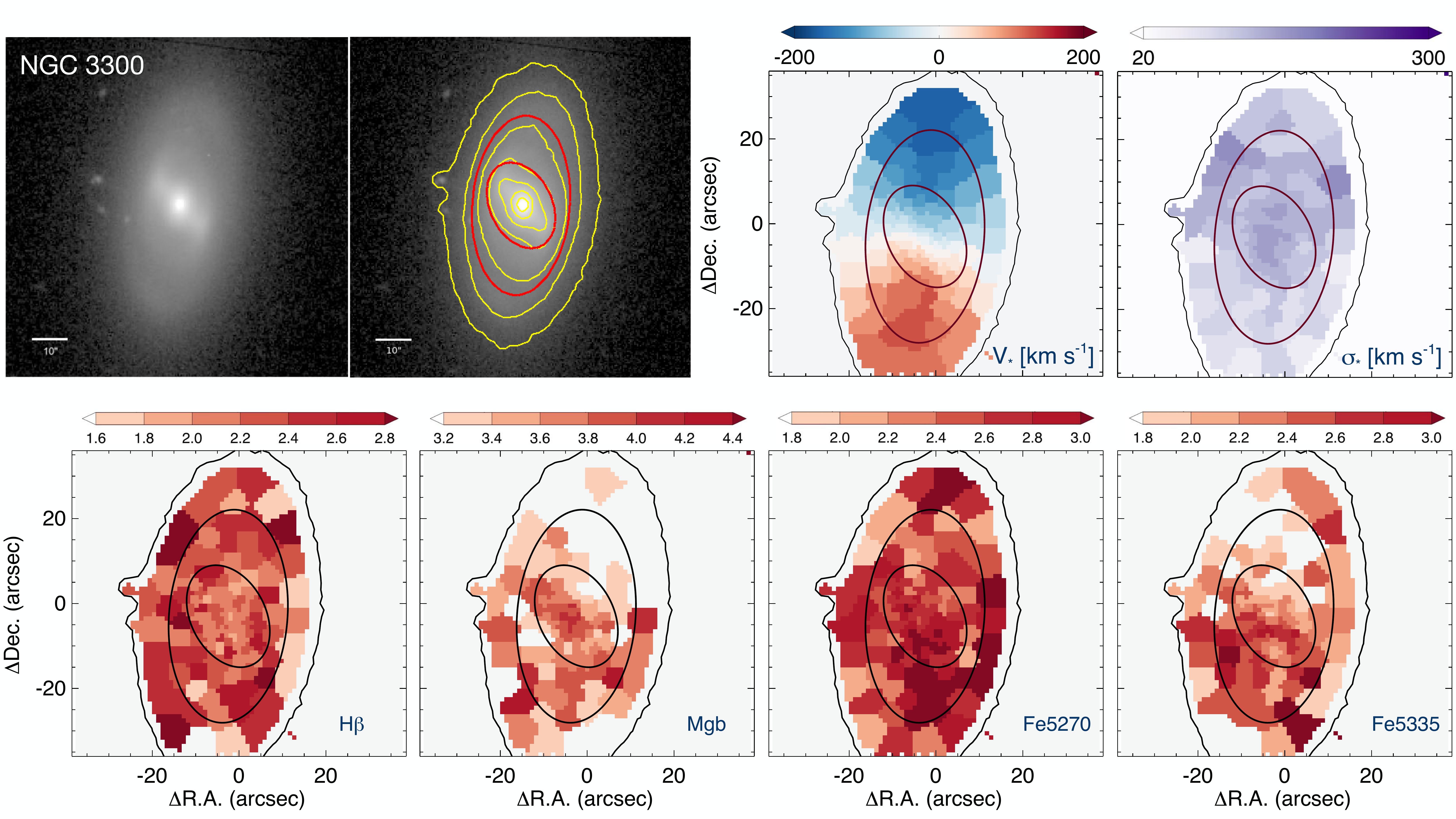}
\caption{ {\it continued}
\label{F45}}
\end{figure*}

\addtocounter{figure}{-1}

\begin{figure*}
\centering
\includegraphics[width=18cm]{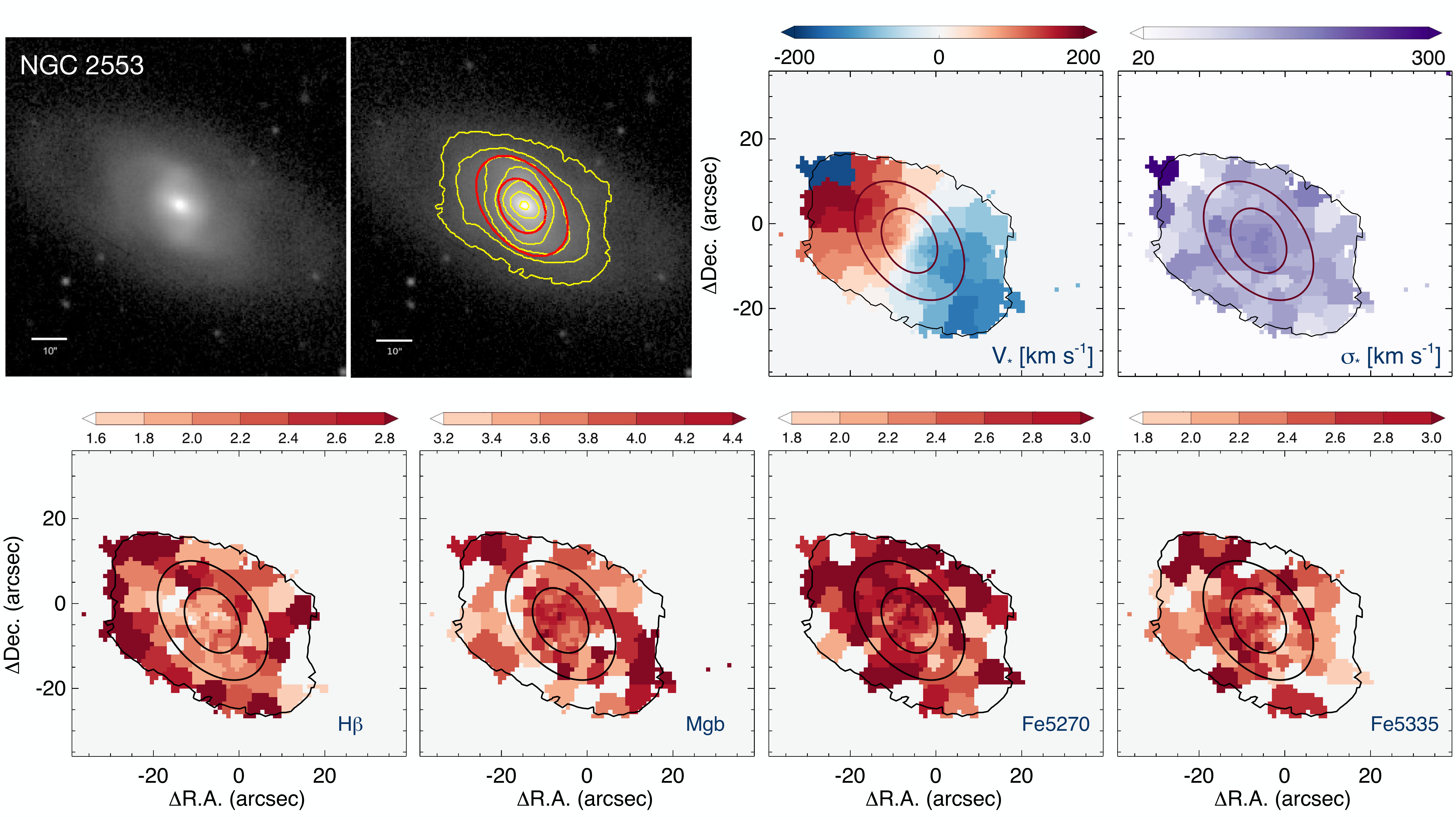}
\caption{ {\it continued}
\label{F46}}
\end{figure*}

\addtocounter{figure}{-1}

\begin{figure*}
\centering
\includegraphics[width=18cm]{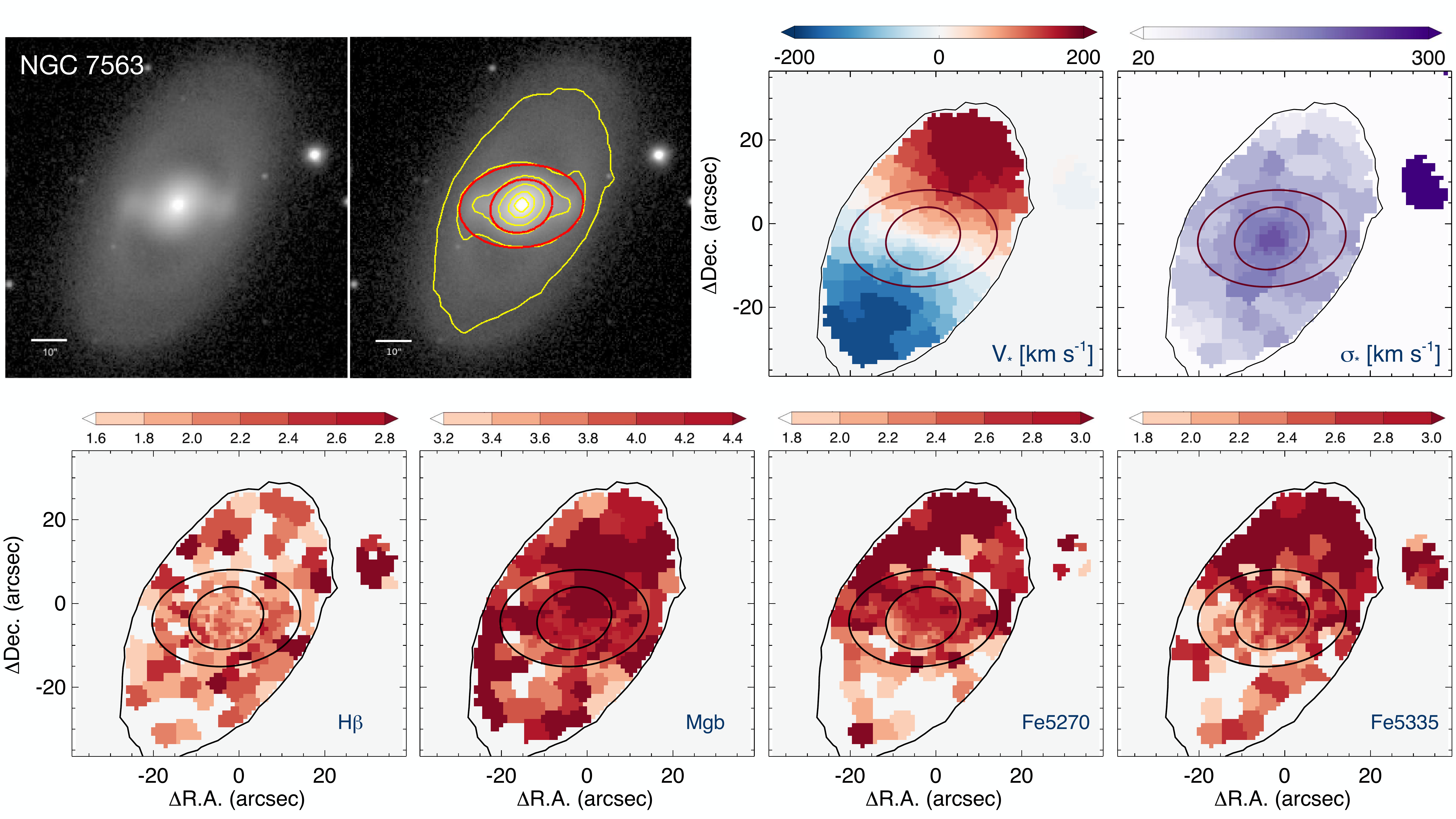}
\caption{ {\it continued}
\label{F47}}
\end{figure*}

\addtocounter{figure}{-1}

\begin{figure*}
\centering
\includegraphics[width=18cm]{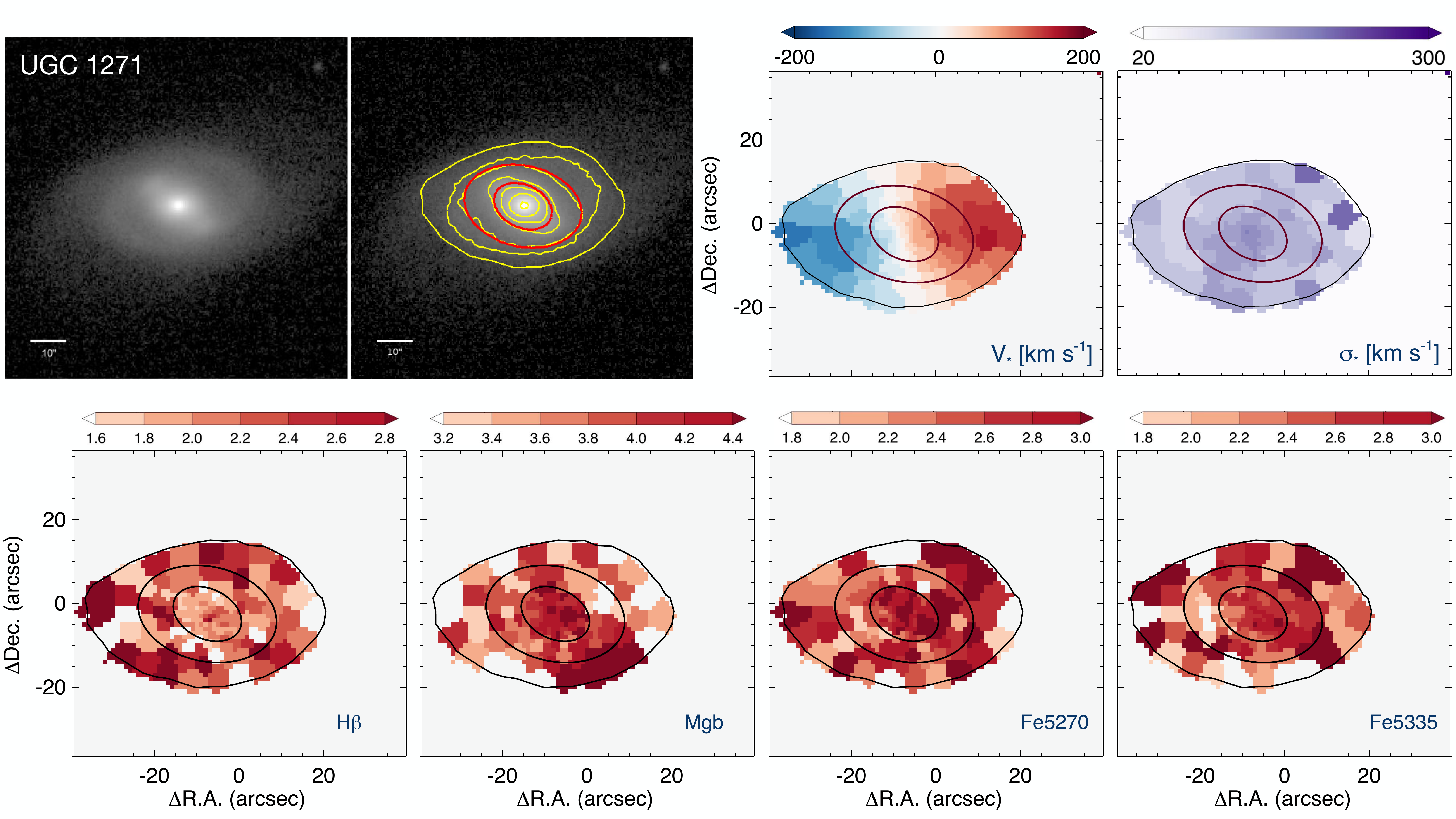}
\caption{ {\it continued}
\label{F48}}
\end{figure*}

\addtocounter{figure}{-1}

\begin{figure*}
\centering
\includegraphics[width=18cm]{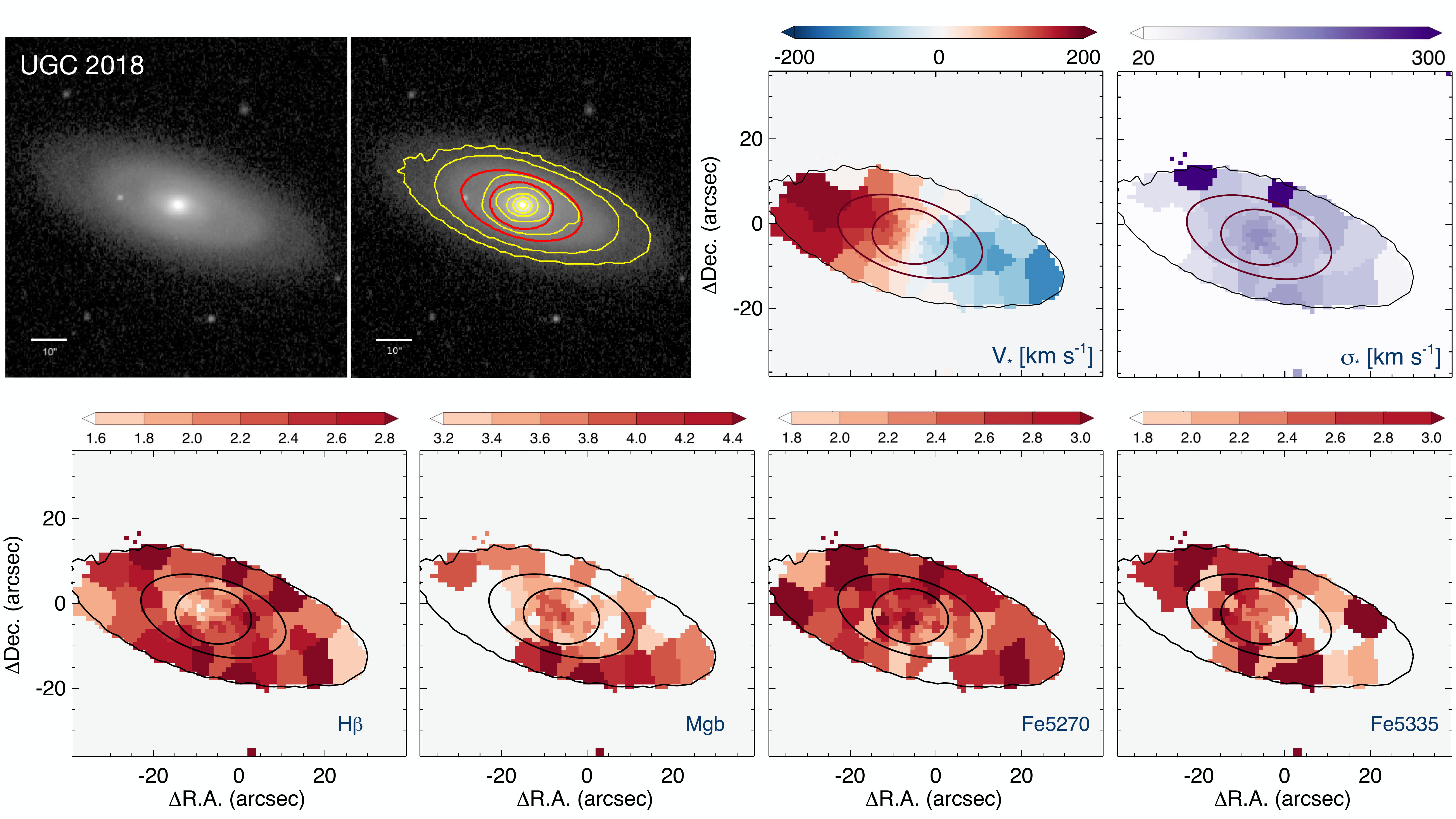}
\caption{ {\it continued}
\label{F49}}
\end{figure*}

\section{Results}
In Figure \ref{F41}, we present the SDSS r-band images, velocity ($v_\star$), velocity dispersion ($\sigma_\star$), and the absorption-line strength maps (e.g. H$\beta$, Mg$b$, Fe5270 and Fe5335) for our passive spiral sample. Our passive spirals have clear rotation and slightly higher velocity dispersion in the center, as also seen in the early type spiral galaxy, NGC 1167 from \citet{Fal17}. We do not find any asymmetry or distorted structures, as evidence for tidal interactions, within their kinematics. In general, H$\beta$ decreases toward the center of the galaxies and the other lines barely show any common trends. Mg$b$ and Fe lines are slightly enhanced along the bar in NGC 3300 and UGC 1271.

\subsection{Stellar populations of passive spiral galaxies}
\begin{figure*}
\includegraphics[width=18cm]{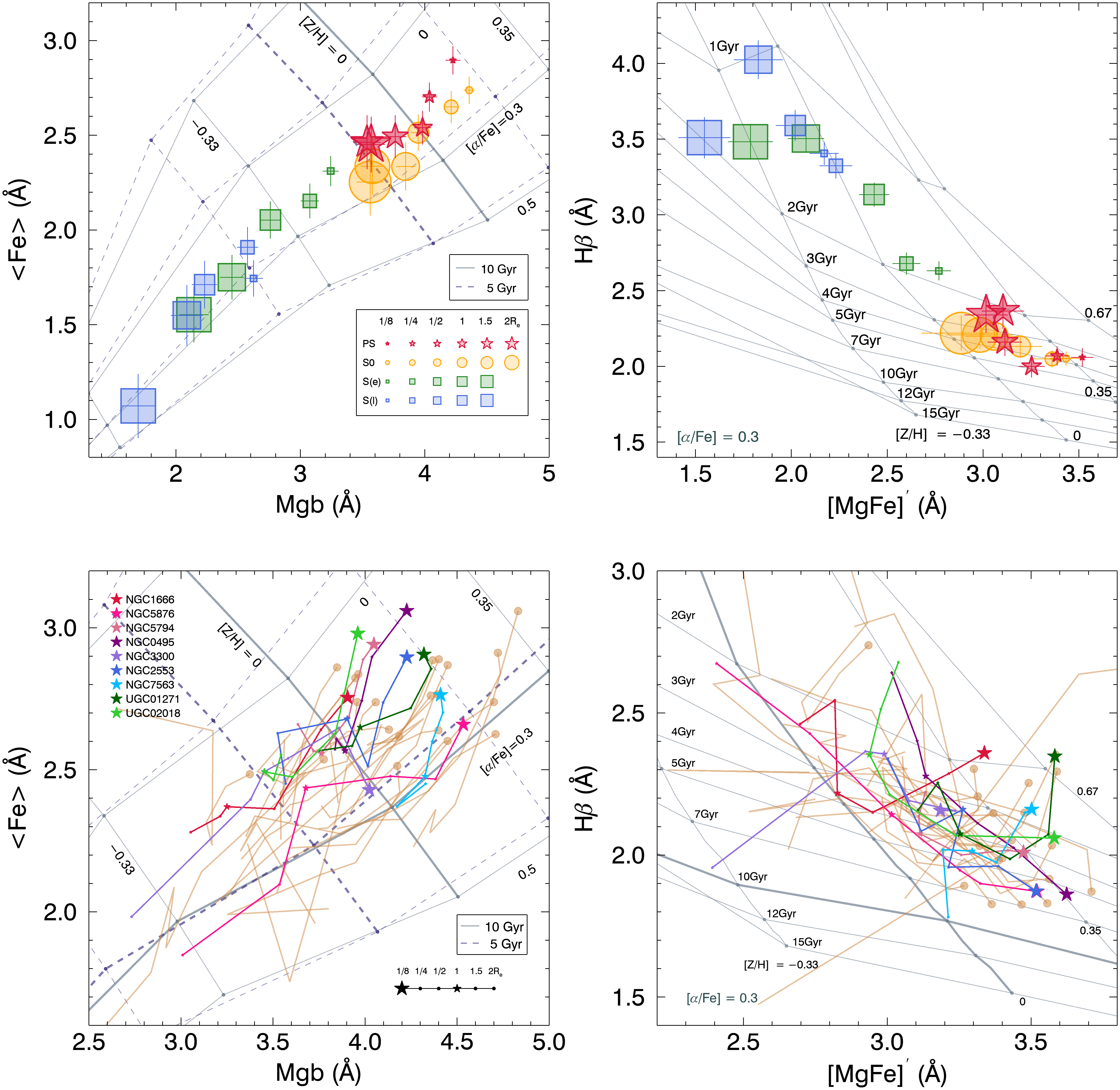}
\caption{Top: Mg$b$ vs. <Fe> (left) and H$\beta$ vs. [MgFe]$^{\prime}$ (right) diagrams for passive spirals (red), spirals (green for Sa-Sb and blue for Sc-Sd spirals) and S0s (yellow) with TMB model grids for ages, metallicities, and abundance ratios. The line strengths are measured along annuli derived from isophotal photometry. The values at $1/8$ R$_e$ (center), $1/4$ R$_e$, $1/2$ R$_e$, $1$ R$_e$, $1.5$ R$_e$ and $2$ R$_e$ are the averages in each annulus with $1 \arcsec$ thickness at the given radius. In the Mg$b$ vs. <Fe> diagram, solid and dashed lines are the $10$ Gyr and $5$ Gyr models, respectively. The model lines of [Z/H] $= 0$ in the left diagram are emphasized as thick lines for each age. Each error bar indicates the median value of measurement uncertainties in each morphological type at the given radius. Bottom: the same index diagrams for passive spirals (colored) and S0s (pale brown) with TMB model grids. The center of each galaxy is marked with large symbols (stars for passive spirals and circles for S0s).
\label{F5}}
\end{figure*}

Figure \ref{F5} presents index-index diagrams with isophotal radial profiles for our sample galaxies from the measurements of the absorption line strengths. For this, we first conduct isophotal photometry for the CALIFA r-band image of each target, using the IRAF/ELLIPSE task \citep{Jed87}. We then use these isophotes to estimate the radial profiles of the line indices.

In the top panels of Figure \ref{F5}, we present the median stellar population properties of each morphological type at each radius in comparison to the TMB model grids. The median line indices were estimated out to $2$ R$_e$, but we note that the signal-to-noise drops below $30$ at R $>$ R$_e$.

In Figure \ref{F5}, [$\alpha$/Fe] appears to be almost constant regardless of galaxy type or radius, whereas metallicity tends to decrease from early- to late-types and from the central to outer regions of galaxies. A common trend is also found for decreasing age and metallicity from early- to late-types, and also with radius. For comparison, we show the star-forming spirals within the same mass range as  the other types out to $1.5$ R$_e$. Why the radial limit is $1.5$ R$_e$, and not $2$ R$_e$, is that most of these spirals have larger R$_e$ than passive spirals and S0s, and the FoV of CALIFA does not reach out to $2$ R$_e$ for more than $30 \%$ of our spirals. The star-forming spiral galaxies are clearly distinct from passive spirals and S0s in both index-index diagrams, which indicates that the stellar populations of passive spirals are closer to those of S0s rather than star-forming spirals. Hence, we focus on the comparison between passive spirals and S0s individually in the bottom panels of Figure \ref{F5}. Overall, we find that our passive spirals show a large variety of behavior among themselves. It is noticeable that S0s have even larger diversity in their stellar populations.

Figure \ref{F6} shows the stellar population maps for our passive spirals. The stellar populations are very diverse as well. NGC 1666 shows overall young ages but is slightly older in the vicinity of $1$ R$_e$. NGC 5876 tends to be older in the central bulge and along the bar. NGC 7563 show mostly old stellar populations. In the metallicity maps, the outer parts tend to be more metal-poor, although the fluctuations are large. [$\alpha$/Fe] tends to be slightly lower at the center, except in NGC 7563.

Figure \ref{F7} shows the distribution of stellar population parameters. Overall, S0s (orange hatched) span a wide range of stellar population parameters at all radii and the distributions of passive spirals (red filled) are encompassed by the ranges of S0 properties for all parameters The probability, P$_0$, of similarity in the cumulative distributions between the two galaxy types as returned by a Kolmogorov-Smirnov (KS) test is given in each panel. However, passive spirals seem to be slightly biased within the S0 distribution: towards higher [Z/H], lower [$\alpha$/Fe], and younger ages. Although such biases do not appear to be very obvious possibly due to the small sample size, some marginal differences are detected in the KS test: P$_0 = 0.13$ for age at $2$ R$_e$, P$_0 = 0.12$ for [Z/H] at $1$ R$_e$, and P$_0 = 0.03$ for [$\alpha$/Fe] at $1$ R$_e$.

One may suspect that this difference at outer region between passive spirals and S0s may be caused by the morphological difference between them, such as the difference in the fraction of bulge light at given radius. However, we confirmed that there is no systematic difference in the distribution of bulge fraction at any radius up to $2$ R$_e$ between our passive spirals and S0s. Both types of galaxies show consistent bulge fraction on average with large dispersion at given radius.

Although the stellar populations of the S0 galaxies are widely dispersed as already mentioned, their metallicities tend to be slightly richer in the center than in the outskirt, whereas their ages seem to be similar between the center and the outskirt. No radial gradient is found for [$\alpha$/Fe] on average. This trend is in good agreement with previous results (\citealt{Tab17}; \citealt{Fra18}).

\begin{figure*}
\centering
\includegraphics[width=18cm]{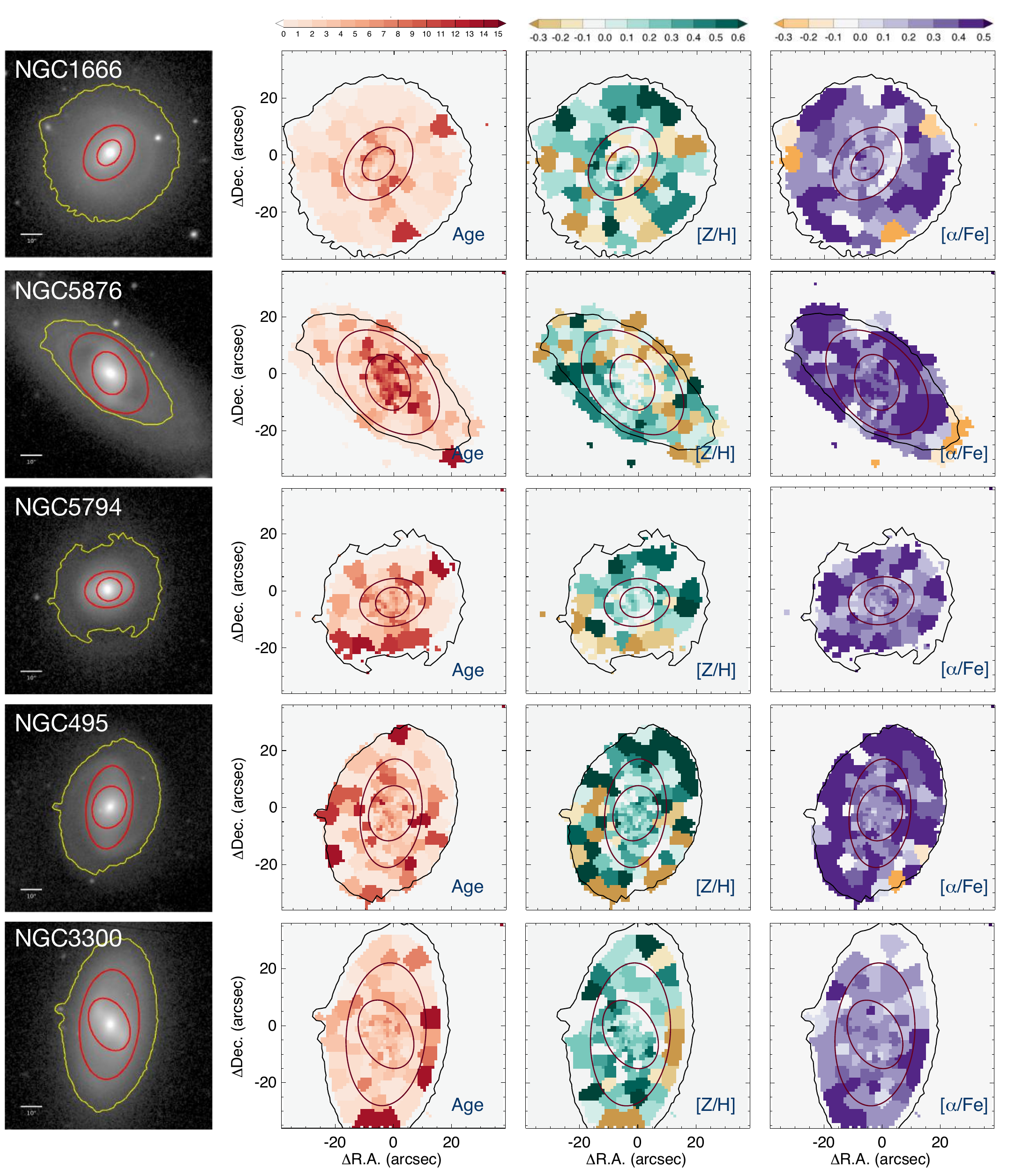}
\caption{SDSS r-band image, derived age, [Z/H], and [$\alpha$/Fe] maps for the passive spirals. The outermost surface brightness contour and the one and two effective radii ellipses are denoted by a yellow line and red ellipses respectively in each r-band image, and by a black line and red ellipses respectively in each stellar population map.
\label{F6}}
\end{figure*}

\addtocounter{figure}{-1}

\begin{figure*}[t]
\centering
\includegraphics[width=18cm]{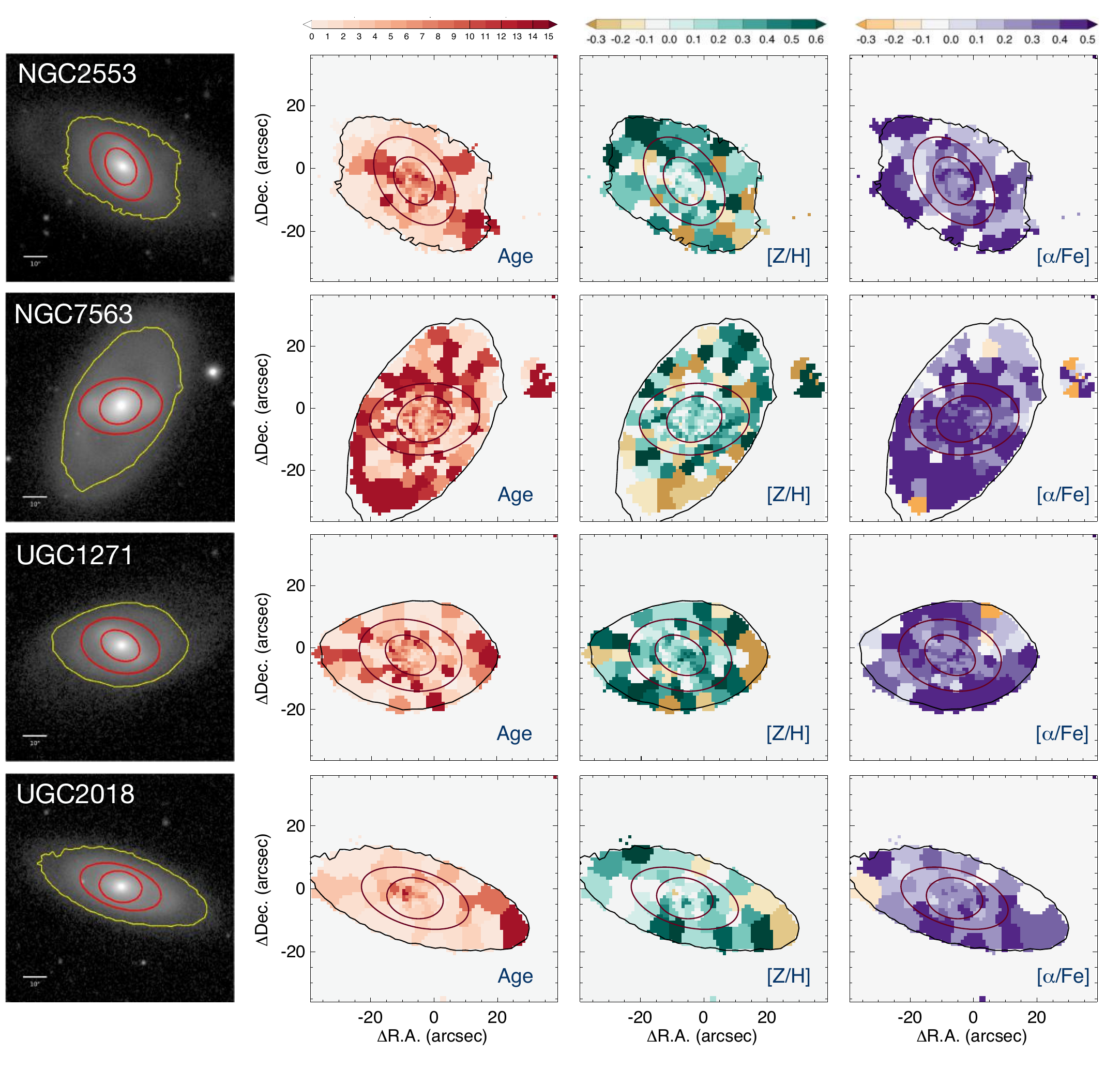}
\caption{ {\it continued}
\label{F62}}
\end{figure*}

\begin{figure}
\includegraphics[width=8.5cm]{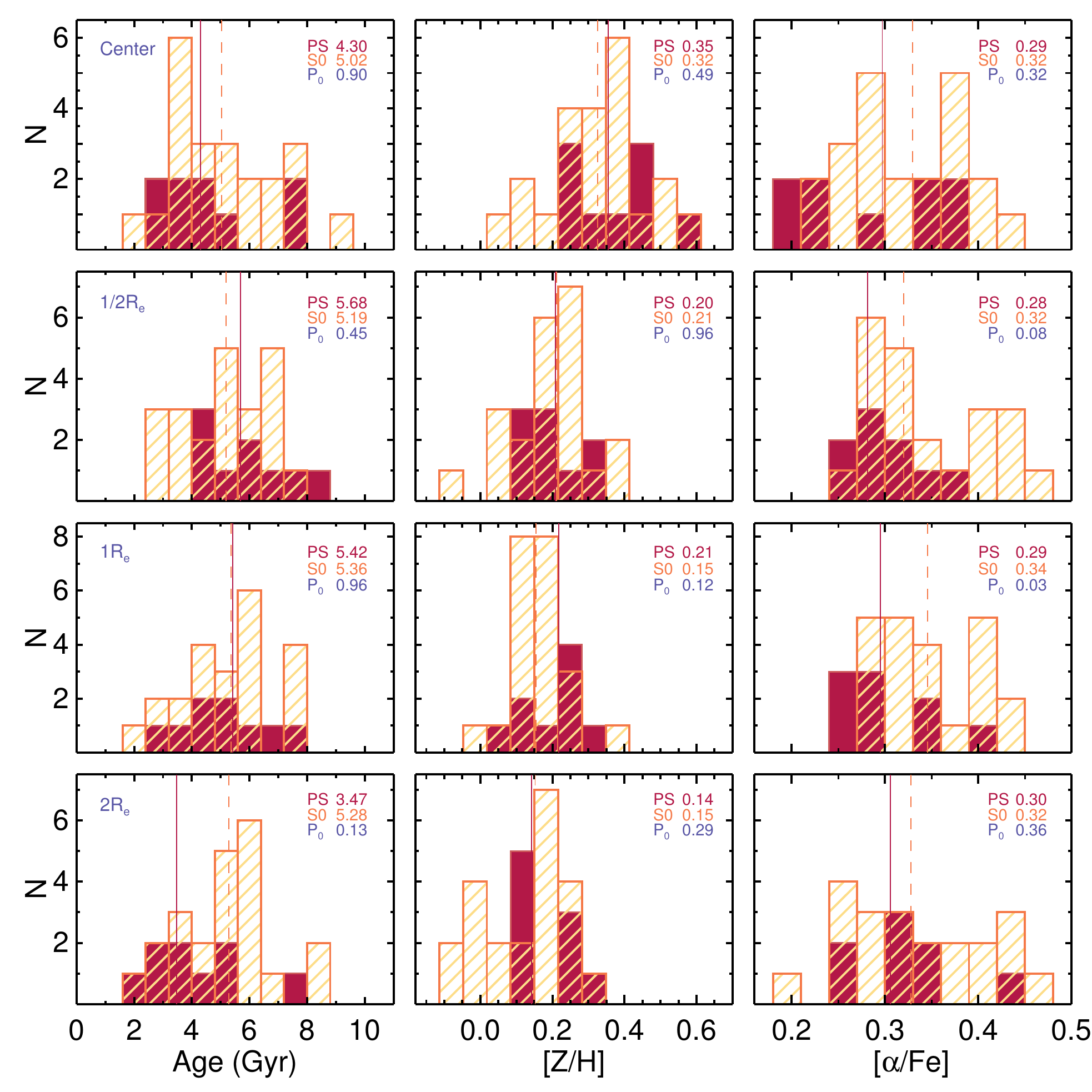}
\caption{Comparison of luminosity-weighted age (left column), metallicity (middle column) and $\alpha$-abundance (right column) at the center ($1/8$ R$_e$), $1/2$ R$_e$, $1$ R$_e$, and $2$ R$_e$ from top to bottom rows for passive spirals (red filled histogram) and S0s (orange hatched histogram). The median values of passive spirals and S0s are denoted in the top right corner of each panel.
\label{F7}}
\end{figure}

\subsection{Environments of passive spiral galaxies}
\begin{figure}[t]
\includegraphics[width=8.5cm]{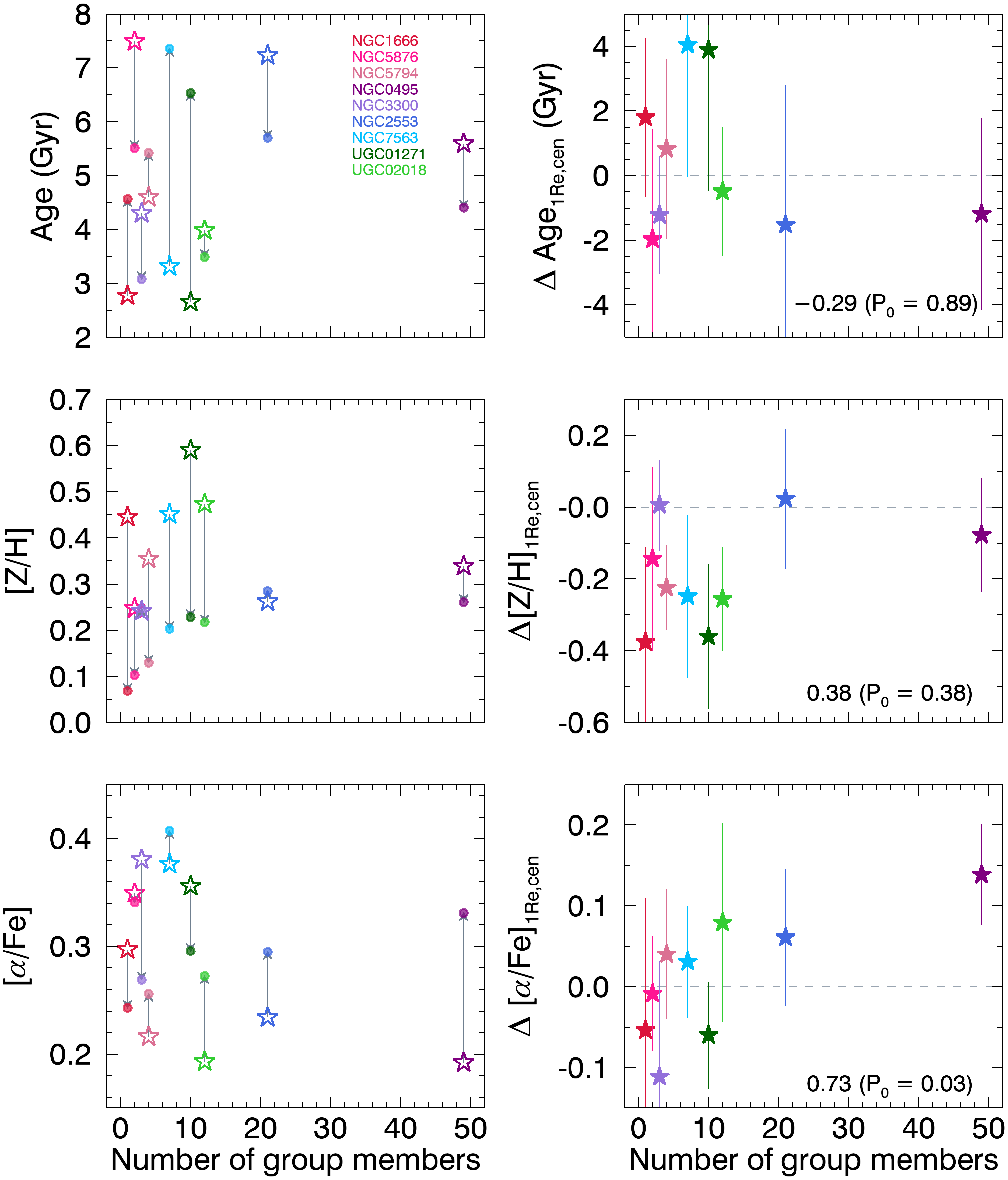}
\caption{Age, metallicity, and $\alpha$-abundance variations between the center (star symbol) and $1$ R$_e$ (dot symbol) in the left column and their differentials ($1$ R$_e$ $-$ center) in the right column for passive spirals with the number of group member. The measurement errors are shown. The Spearman's correlation coefficient and probability (P$_0$) are presented in the right panels. We adopt the membership information of each galaxy from the group catalog of \citet{Tul15}.
\label{F8}}
\end{figure}

Next, we investigate how the stellar populations of passive spirals vary in each individual galaxy with their environment in Figure \ref{F8}. We adopt the number of group members (N$_m$) as an environmental indicator by using the group catalog of \citet{Tul15}. According to the information in this group catalog, our passive spiral sample exists over all environments from field to cluster. For instance, NGC 1666 is in isolation, while NGC 495 is a part of the Perseus-Pisces Supercluster. 

The results show that the correlations are mostly unclear between the stellar populations and environment of the passive spirals in our sample. The $\Delta$age and $\Delta$[Z/H]] do not exhibit significant correlations with environments: Spearman's rank coefficient is $-0.29$ (the significance of deviation, P$_0 = 0.89$) for the age and $0.38$ (P$_0 = 0.38$) for the metallicity. On the other hand, even though there is the caveat of small-number statistics, we obtain a Spearman's rank coefficient for $\Delta$[$\alpha$/Fe] of $0.73$ (P$_0 = 0.03$). However, when we eliminate the single point in the upper-right corner (NGC 495) which makes the correlation look strong, the correlation coefficient becomes smaller ($0.59$ with P$_0 = 0.12$) but still is considerable. This means that, in denser environments, at 1 R$_e$ passive spirals have higher [$\alpha$/Fe] than in their centers, and this likely indicates they were quenched faster in their outskirt than in their centers, perhaps by physical mechanisms operating in the denser environment. Even though there is no correlation between [Z/H] and environment, the gradient mostly show negative values.

\section{Discussion and Conclusions}
Now we discuss the evolutionary pathways for passive spirals. Today, one of the widely-believed origins of S0s is that they evolved from spiral galaxies. That is, S0s may be the end-product of the transformation of spiral galaxies that have had their gas content removed by means of external gravitational or non-gravitational forces, or by secular evolution. In this scenario, passive spirals could be objects caught in the transitional stage between normal star-forming spirals and S0s \citep{Bek02}. Previous studies argued that multiple quenching mechanisms are required (\citealt{Mas10}; \citealt{Fra18}). Our results are consistent with this suggestion to some extent. 

There seem to exist evolutionary connections between passive spirals and S0s from the comparison of their stellar populations. The passive spirals may be the origin of at least a portion of S0s, with other evolutionary pathways contributing in parallel. By comparing stellar populations in Figs \ref{F5} and \ref{F7}, it is revealed that the S0s have a wider range of age, metallicity and $\alpha$-abundance at all radii out to $2$ R$_e$. The stellar populations of passive spirals are fully encompassed within the spread of S0 properties, although the distributions of passive spirals appear skewed toward slightly younger ages, higher metallicities, and lower $\alpha$-abundances.

For the origin of passive spirals, quenching by perturbation from a bar is one plausible scenario. As one can see in Fig. \ref{F1}, $7$ out of the $9$ passive spirals have clear bar features. This high bar fraction ($\sim 78 \%$) in passive spirals is consistent with previous results. \citet{Mas10} reported that red spirals have an optical bar fraction of at least $67 \pm 5\%$. On the other hand, \citet{Fra18} found that the bar fraction of passive spiral galaxies is significantly higher ($74 \pm 15\%$), compared to mass, redshift and T-type matched star-forming spiral galaxies ($36 \pm 5\%$). Therefore, we suspect that bars in galaxies, both in low and high density environments, may act to shut down the star formation, leaving behind spiral structures that may last for several Gyrs \citep{Bek02}. However, it is unclear if secular evolution by a bar is the only mechanism for quenching our passive spiral sample. Even though a large fraction of our passive spirals have bars in their disks, there still exist non-barred ones which could only be explained in this scenario if bars are transient phenomena \citep{Bou02}.

Quenching by environmental effects is also a possible scenario. As one can see in our results in Figure \ref{F8}, the $\Delta[\alpha$/Fe] of passive spirals are marginally related with their environments. It means that environmental effects may have influenced the quenching in outskirt of a galaxy. All our passive spirals, except NGC 1666 in isolation, have close neighbors ($\sim 47 - 274$ kpc), so there are opportunities for galaxies to be affected by their neighbors, even in sparse groups. 

These passive spirals may have already depleted all their gas reservoirs, or they still have plenty of gas but it is not condensed enough to trigger star formation. Recently, \citet{Fre18} found a significant quantity of CO-traced low density gas in two (one is highly disturbed) post-starbursts (or `E+A') galaxies, which are currently quiescent galaxies. Similarly, \citep{Ell18} found there is still significant quantities of gas in some galaxy-galaxy mergers even after star formation was halted. Therefore, we check the gas contents of our passive spirals case by case. Table \ref{Tbl2} shows a summary of the information from literature about the intracluster medium (ICM) in their host groups, and HI and CO observations in each galaxy. We confirm that, except for NGC 5794 and NGC 495 (which were not observed in the HI Parkes All Sky Survey; HIPASS), our passive spirals were targeted but not detected. Therefore, based on the detection limit of the HIPASS ($10^{6} \times d_{Mpc}^{2}$; \citealt{Sta00}), we can place upper-limit on their HI masses of $\leq$ $ 10^{8.9} - 10^{9.6}$ M$_{\odot}$. This amount is obviously lower than the median HI masses of normal spiral galaxies in the HIPASS; log(M$_{HI}$/M$_{\odot}$) $= 9.59 \pm 0.43$ and $9.61 \pm 0.46$ in the stellar masses of log(M${_\star}$/M${_\odot}) = 10.25$ and $10.75$, respectively (\citealt{Par18}). Here, we describe some noticeable cases as follows. \\

\begin{table}   
\caption{X-ray, HI and CO information of $9$ passive spirals}
\centering
 \label{Tbl2}
\begin{tabular}{llrrcccccc}
\hline
Name     & Group    &   N$_m$  & D$_{1st}$   & HI        & CO        & X-ray    \\
         &          &          & (kpc)       &           &           &          \\
(1)      & (2)      & (3)      & (4)         & (5)       & (6)       & (7)    \\
\hline
NGC 1666 & NGC 1666 &  1       &  -          & $\times$  & -         & $\times$ \\
NGC 5876 & NGC 5876 &  2       & 248         & -         & $\times$  &  -       \\
NGC 5794 & NGC 5797 &  4       &  78         & -         & -         &  -       \\
NGC 495  & NGC 507  &  49      &  47         & -         & -         & $\circ$  \\
NGC 3300 & NGC 3300 &  3       & 274         & $\times$  & -         &  -       \\
NGC 2553 & NGC 2563 &  21      & 229         & $\times$  & -         & $\circ$  \\
NGC 7563 & NGC 7563 &  7       &  76         & $\times$  & -         &  -       \\
UGC 1271 & NGC 677  &  10      & 132         & $\times$  & -         & $\circ$  \\
UGC 2018 & UGC 2005 &  12      & 203         & $\times$  & -         &  -       \\
\hline
\end{tabular}
\\
\begin{flushleft}
Notes. Column 1 gives the name of passive spirals, and Column 2 to 4 give the host galaxy of the group to which each passive spiral belongs, number of members (N$_m$), and the distance from the nearest neighbor adopted from \citet{Tul15}, respectively. Column 4 and 5 show the HI (HIPASS) and CO (EDGE-CALIFA Survey from \citealt{Bol17}) information of each galaxy. Column 6 describes if the group has an X-ray source): NGC 495 (\citealt{kim04}; \citealt{Sat09}), NGC 2553 (\citealt{Ras12}; \citealt{Mor17}) and UGC 1271 (\citealt{Osu17}). The symbols of `$\circ$' and `$\times$' denote if the source is `detected' or `non-detected', respectively. If the observation does not exist, we mark it with ` - '.
\end{flushleft}
\end{table}
\noindent \textit{NGC 5876.} NGC 5876 is listed in the recent CO observation of the EDGE-CALIFA Survey \citep{Bol17}, but no significant emission is detected (S$_{CO}$ $\Delta$v (Jy km s$^{â1}$) $<16.0$). This galaxy is in a pair with a normal spiral galaxy at a distance of $248$ kpc but there is no evidence for a direct interaction in the optical image. \\

\noindent \textit{NGC 2553.} Although NGC 2553 is quite far from the center of its group, no meaningful signal of HI is detected in this galaxy. The ICM is detected in this group (\citealt{Ras12}; \citealt{Mor17}). Therefore, this galaxy may be influenced by the environmental effects such as ram-pressure stripping or harassment. \\

\noindent \textit{NGC 1666.} The isolated galaxy NGC 1666 is notable. There is no meaningful HI signal in this galaxy, despite the fact that it may have rarely experienced environmental effects. In this case, secular evolution may be the main driver for star formation quenching. If there is no further inflow of fresh gas, the gas in the disk will run out in a few Gyrs (starvation or strangulation; \citealt{Lar80}). Especially, in the center of the galaxy, the age is the youngest and the metallicity is the richest among all of our passive spiral sample (see Fig. \ref{F8}). Hence, this galaxy may be quenched very recently. However, the question of how all its gas is used up by itself still remains. The enrichment of metallicity is also not easily understood but it may imply that the inflow of pristine gas has been cut off for some reason. \\

\noindent \textit{NGC 495.} NGC 495 is in the Perseus-Pisces Supercluster. This galaxy is close to the central galaxy of the cluster ($\sim200$ kpc) and has a close early-type neighbor at a distance of $\sim47$ kpc. In addition, there is ICM detected in this group (\citealt{kim04}; \citealt{Sat09}), so NGC 495 may have been influenced by ram-pressure stripping. Thus, we might expect that there is no or a very small amount of cold gas in this galaxy, but unfortunately NGC 495 is out of the coverage of the HIPASS. \\

We conclude that the similarity in stellar populations between passive spirals and S0s is in accordance with the idea that S0s may have formed through various evolutionary pathways, and passive spirals could be one of the channels transforming from spirals to S0s. (1) Passive spirals in isolation or low density regions might be just old spirals that have exhausted their fuel. Such gas consumption may have been accelerated by internal processes related with bar structure. (2) In denser environments, neighbor interactions or group/cluster mechanisms can help strip the gas away predominantly in the halo and outer disks, but still secular evolution by a bar could accelerate quenching. Both scenarios may explain the cessation of star forming activity without destroying the spiral structure.

The analysis of large public data sets such as those of the Sydney-AAO Multi-object Integral field Spectrograph galaxy survey (SAMI; \citealt{Cro12}) and the Mapping Nearby Galaxies at APO (MaNGA; \citealt{Bun15}) would overcome the small number statistics in this work, and provide deeper insights on their detailed evolutionary pathways.

\section*{Acknowledgments}

We gratefully thank the anonymous referee for constructive comments that have significantly improved this manuscript. This study uses data provided by the Calar Alto Legacy Integral Field Area (CALIFA) survey (http://califa.caha.es/). Based on observations collected at the Centro Astron\'omico Hispano Alem\'an (CAHA) at Calar Alto, operated jointly by the Max-Planck-Institut f\"ur Astronomie and the Instituto de Astrof\'isica de Andaluc\'ia (CSIC). H.J. acknowledges support from the Basic Science Research Program through the National Research Foundation (NRF) of Korea, funded by the Ministry of Education (NRF-2013R1A6A3A04064993).


\begin{thebibliography}{}

\bibitem[{{Athanassoula}(2003)}]{Ath03}{Athanassoula}, E. 2003, \mnras, 341, 1179

\bibitem[{{Athanassoula}(2013)}]{Ath13}---. 2013, {Bars and secular evolution in disk galaxies: Theoretical input},  ed. J.~{Falc{\'o}n-Barroso} \& J.~H. {Knapen}, 305

\bibitem[{{Bamford} {et~al.}(2009){Bamford}, {Nichol}, {Baldry}, {Land},  {Lintott}, {Schawinski}, {Slosar}, {Szalay}, {Thomas}, {Torki}, {Andreescu},  {Edmondson}, {Miller}, {Murray}, {Raddick}, \& {Vandenberg}}]{Bam09}{Bamford}, S.~P., {Nichol}, R.~C., {Baldry}, I.~K., {et~al.} 2009, \mnras, 393,  1324

\bibitem[{{Barazza} {et~al.}(2008){Barazza}, {Jogee}, \& {Marinova}}]{Bar08}
{Barazza}, F.~D., {Jogee}, S., \& {Marinova}, I. 2008, \apj, 675, 1194

\bibitem[{{Bekki} {et~al.}(2002){Bekki}, {Couch}, \& {Shioya}}]{Bek02}
{Bekki}, K., {Couch}, W.~J., \& {Shioya}, Y. 2002, \apj, 577, 651

\bibitem[{{Bershady} {et~al.}(2000){Bershady}, {Jangren}, \&
  {Conselice}}]{Ber00}
{Bershady}, M.~A., {Jangren}, A., \& {Conselice}, C.~J. 2000, \aj, 119, 2645

\bibitem[{{Blanton} {et~al.}(2011){Blanton}, {Kazin}, {Muna}, {Weaver}, \&
  {Price-Whelan}}]{Bla11}
{Blanton}, M.~R., {Kazin}, E., {Muna}, D., {Weaver}, B.~A., \& {Price-Whelan},
  A. 2011, \aj, 142, 31

\bibitem[{{Bolatto} {et~al.}(2017){Bolatto}, {Wong}, {Utomo}, {Blitz}, {Vogel},
  {S{\'a}nchez}, {Barrera-Ballesteros}, {Cao}, {Colombo}, {Dannerbauer},
  {Garc{\'{\i}}a-Benito}, {Herrera-Camus}, {Husemann}, {Kalinova}, {Leroy},
  {Leung}, {Levy}, {Mast}, {Ostriker}, {Rosolowsky}, {Sandstrom}, {Teuben},
  {van de Ven}, \& {Walter}}]{Bol17}
{Bolatto}, A.~D., {Wong}, T., {Utomo}, D., {et~al.} 2017, \apj, 846, 159

\bibitem[{{Bournaud} \& {Combes}(2002)}]{Bou02}
{Bournaud}, F., \& {Combes}, F. 2002, \aap, 392, 83

\bibitem[{{Bundy} {et~al.}(2015){Bundy}, {Bershady}, {Law}, {Yan}, {Drory},
  {MacDonald}, {Wake}, {Cherinka}, {S{\'a}nchez-Gallego}, {Weijmans}, {Thomas},
  {Tremonti}, {Masters}, {Coccato}, {Diamond-Stanic}, {Arag{\'o}n-Salamanca},
  {Avila-Reese}, {Badenes}, {Falc{\'o}n-Barroso}, {Belfiore}, {Bizyaev},
  {Blanc}, {Bland-Hawthorn}, {Blanton}, {Brownstein}, {Byler}, {Cappellari},
  {Conroy}, {Dutton}, {Emsellem}, {Etherington}, {Frinchaboy}, {Fu}, {Gunn},
  {Harding}, {Johnston}, {Kauffmann}, {Kinemuchi}, {Klaene}, {Knapen},
  {Leauthaud}, {Li}, {Lin}, {Maiolino}, {Malanushenko}, {Malanushenko}, {Mao},
  {Maraston}, {McDermid}, {Merrifield}, {Nichol}, {Oravetz}, {Pan}, {Parejko},
  {Sanchez}, {Schlegel}, {Simmons}, {Steele}, {Steinmetz}, {Thanjavur},
  {Thompson}, {Tinker}, {van den Bosch}, {Westfall}, {Wilkinson}, {Wright},
  {Xiao}, \& {Zhang}}]{Bun15}
{Bundy}, K., {Bershady}, M.~A., {Law}, D.~R., {et~al.} 2015, \apj, 798, 7

\bibitem[{{Buta}(2013)}]{But13}
{Buta}, R.~J. 2013, {Galaxy Morphology}, ed. J.~{Falc{\'o}n-Barroso} \& J.~H.
  {Knapen}, 155

\bibitem[{{Cappellari} \& {Copin}(2003)}]{Cap03}
{Cappellari}, M., \& {Copin}, Y. 2003, \mnras, 342, 345

\bibitem[{{Cortese} \& {Hughes}(2009)}]{Cor09}
{Cortese}, L., \& {Hughes}, T.~M. 2009, \mnras, 400, 1225

\bibitem[{{Couch} {et~al.}(1998){Couch}, {Barger}, {Smail}, {Ellis}, \&
  {Sharples}}]{Cou98}
{Couch}, W.~J., {Barger}, A.~J., {Smail}, I., {Ellis}, R.~S., \& {Sharples},
  R.~M. 1998, \apj, 497, 188

\bibitem[{{Cowie} \& {Songaila}(1977)}]{Cow77}
{Cowie}, L.~L., \& {Songaila}, A. 1977, \nat, 266, 501

\bibitem[{{Croom} {et~al.}(2012){Croom}, {Lawrence}, {Bland-Hawthorn},
  {Bryant}, {Fogarty}, {Richards}, {Goodwin}, {Farrell}, {Miziarski}, {Heald},
  {Jones}, {Lee}, {Colless}, {Brough}, {Hopkins}, {Bauer}, {Birchall}, {Ellis},
  {Horton}, {Leon-Saval}, {Lewis}, {L{\'o}pez-S{\'a}nchez}, {Min}, {Trinh}, \&
  {Trowland}}]{Cro12}
{Croom}, S.~M., {Lawrence}, J.~S., {Bland-Hawthorn}, J., {et~al.} 2012, \mnras,
  421, 872

\bibitem[{{Debattista} \& {Sellwood}(1998)}]{Deb98}
{Debattista}, V.~P., \& {Sellwood}, J.~A. 1998, \apjl, 493, L5

\bibitem[{{Debattista} \& {Sellwood}(2000)}]{Deb00}
---. 2000, \apj, 543, 704

\bibitem[{{Dressler} {et~al.}(1999){Dressler}, {Smail}, {Poggianti}, {Butcher},
  {Couch}, {Ellis}, \& {Oemler}}]{Dre99}
{Dressler}, A., {Smail}, I., {Poggianti}, B.~M., {et~al.} 1999, \apjs, 122, 51

\bibitem[{{Ellison} {et~al.}(2018){Ellison}, {Catinella}, \& {Cortese}}]{Ell18}
{Ellison}, S.~L., {Catinella}, B., \& {Cortese}, L. 2018, \mnras, 478, 3447

\bibitem[{{Erwin}(2004)}]{Erw04}
{Erwin}, P. 2004, \aap, 415, 941

\bibitem[{{Falc{\'o}n-Barroso} {et~al.}(2017){Falc{\'o}n-Barroso}, {Lyubenova},
  {van de Ven}, {Mendez-Abreu}, {Aguerri}, {Garc{\'{\i}}a-Lorenzo},
  {Bekerait{\'e}}, {S{\'a}nchez}, {Husemann}, {Garc{\'{\i}}a-Benito}, {Mast},
  {Walcher}, {Zibetti}, {Barrera-Ballesteros}, {Galbany},
  {S{\'a}nchez-Bl{\'a}zquez}, {Singh}, {van den Bosch}, {Wild}, {Zhu},
  {Bland-Hawthorn}, {Cid Fernandes}, {de Lorenzo-C{\'a}ceres}, {Gallazzi},
  {Gonz{\'a}lez Delgado}, {Marino}, {M{\'a}rquez}, {P{\'e}rez}, {P{\'e}rez},
  {Roth}, {Rosales-Ortega}, {Ruiz-Lara}, {Wisotzki}, {Ziegler}, \& {Califa
  Collaboration}}]{Fal17}
{Falc{\'o}n-Barroso}, J., {Lyubenova}, M., {van de Ven}, G., {et~al.} 2017,
  \aap, 597, A48

\bibitem[{{Fioc} \& {Rocca-Volmerange}(1999)}]{Fio99}
{Fioc}, M., \& {Rocca-Volmerange}, B. 1999, \aap, 351, 869

\bibitem[{{Fraser-McKelvie} {et~al.}(2018){Fraser-McKelvie}, {Brown},
  {Pimbblet}, {Dolley}, \& {Bonne}}]{Fra18}
{Fraser-McKelvie}, A., {Brown}, M.~J.~I., {Pimbblet}, K., {Dolley}, T., \&
  {Bonne}, N.~J. 2018, \mnras, 474, 1909

\bibitem[{{Fraser-McKelvie} {et~al.}(2016){Fraser-McKelvie}, {Brown},
  {Pimbblet}, {Dolley}, {Crossett}, \& {Bonne}}]{Fra16}
{Fraser-McKelvie}, A., {Brown}, M.~J.~I., {Pimbblet}, K.~A., {et~al.} 2016,
  \mnras, 462, L11

\bibitem[{{French} {et~al.}(2018){French}, {Zabludoff}, {Yoon}, {Shirley},
  {Yang}, {Smercina}, {Smith}, \& {Narayanan}}]{Fre18}
{French}, K.~D., {Zabludoff}, A.~I., {Yoon}, I., {et~al.} 2018, \apj, 861, 123

\bibitem[{{Garc{\'{\i}}a-Benito} {et~al.}(2015){Garc{\'{\i}}a-Benito},
  {Zibetti}, {S{\'a}nchez}, {Husemann}, {de Amorim}, {Castillo-Morales}, {Cid
  Fernandes}, {Ellis}, {Falc{\'o}n-Barroso}, {Galbany}, {Gil de Paz},
  {Gonz{\'a}lez Delgado}, {Lacerda}, {L{\'o}pez-Fernandez}, {de
  Lorenzo-C{\'a}ceres}, {Lyubenova}, {Marino}, {Mast}, {Mendoza}, {P{\'e}rez},
  {Vale Asari}, {Aguerri}, {Ascasibar}, {Bekerait{\.e}}, {Bland-Hawthorn},
  {Barrera-Ballesteros}, {Bomans}, {Cano-D{\'{\i}}az},
  {Catal{\'a}n-Torrecilla}, {Cortijo}, {Delgado-Inglada}, {Demleitner},
  {Dettmar}, {D{\'{\i}}az}, {Florido}, {Gallazzi}, {Garc{\'{\i}}a-Lorenzo},
  {Gomes}, {Holmes}, {Iglesias-P{\'a}ramo}, {Jahnke}, {Kalinova}, {Kehrig},
  {Kennicutt}, {L{\'o}pez-S{\'a}nchez}, {M{\'a}rquez}, {Masegosa}, {Meidt},
  {Mendez-Abreu}, {Moll{\'a}}, {Monreal-Ibero}, {Morisset}, {del Olmo},
  {Papaderos}, {P{\'e}rez}, {Quirrenbach}, {Rosales-Ortega}, {Roth},
  {Ruiz-Lara}, {S{\'a}nchez-Bl{\'a}zquez}, {S{\'a}nchez-Menguiano}, {Singh},
  {Spekkens}, {Stanishev}, {Torres-Papaqui}, {van de Ven}, {Vilchez},
  {Walcher}, {Wild}, {Wisotzki}, {Ziegler}, {Alves}, {Barrado}, {Quintana}, \&
  {Aceituno}}]{Gar15}
{Garc{\'{\i}}a-Benito}, R., {Zibetti}, S., {S{\'a}nchez}, S.~F., {et~al.} 2015,
  \aap, 576, A135

\bibitem[{{Goto} {et~al.}(2003){Goto}, {Okamura}, {Sekiguchi}, {Bernardi},
  {Brinkmann}, {G{\'o}mez}, {Harvanek}, {Kleinman}, {Krzesinski}, {Long},
  {Loveday}, {Miller}, {Neilsen}, {Newman}, {Nitta}, {Sheth}, {Snedden}, \&
  {Yamauchi}}]{Got03}
{Goto}, T., {Okamura}, S., {Sekiguchi}, M., {et~al.} 2003, \pasj, 55, 757

\bibitem[{{Graves} \& {Schiavon}(2008)}]{Gra08}
{Graves}, G.~J., \& {Schiavon}, R.~P. 2008, \apjs, 177, 446

\bibitem[{{Gunn} \& {Gott}(1972)}]{Gun72}
{Gunn}, J.~E., \& {Gott}, III, J.~R. 1972, \apj, 176, 1

\bibitem[{{Hawarden} {et~al.}(1986){Hawarden}, {Mountain}, {Leggett}, \&
  {Puxley}}]{Haw86}
{Hawarden}, T.~G., {Mountain}, C.~M., {Leggett}, S.~K., \& {Puxley}, P.~J.
  1986, \mnras, 221, 41P

\bibitem[{{Hubble}(1938)}]{Hub38}
{Hubble}, E. 1938, \pasp, 50, 97

\bibitem[{{Hughes} \& {Cortese}(2009)}]{Hug09}
{Hughes}, T.~M., \& {Cortese}, L. 2009, \mnras, 396, L41

\bibitem[{{Husemann} {et~al.}(2013){Husemann}, {Jahnke}, {S{\'a}nchez},
  {Barrado}, {Bekerait{\.e}}, {Bomans}, {Castillo-Morales},
  {Catal{\'a}n-Torrecilla}, {Cid Fernandes}, {Falc{\'o}n-Barroso},
  {Garc{\'{\i}}a-Benito}, {Gonz{\'a}lez Delgado}, {Iglesias-P{\'a}ramo},
  {Johnson}, {Kupko}, {L{\'o}pez-Fernandez}, {Lyubenova}, {Marino}, {Mast},
  {Miskolczi}, {Monreal-Ibero}, {Gil de Paz}, {P{\'e}rez}, {P{\'e}rez},
  {Rosales-Ortega}, {Ruiz-Lara}, {Schilling}, {van de Ven}, {Walcher}, {Alves},
  {de Amorim}, {Backsmann}, {Barrera-Ballesteros}, {Bland-Hawthorn}, {Cortijo},
  {Dettmar}, {Demleitner}, {D{\'{\i}}az}, {Enke}, {Florido}, {Flores},
  {Galbany}, {Gallazzi}, {Garc{\'{\i}}a-Lorenzo}, {Gomes}, {Gruel}, {Haines},
  {Holmes}, {Jungwiert}, {Kalinova}, {Kehrig}, {Kennicutt}, {Klar}, {Lehnert},
  {L{\'o}pez-S{\'a}nchez}, {de Lorenzo-C{\'a}ceres}, {M{\'a}rmol-Queralt{\'o}},
  {M{\'a}rquez}, {Mendez-Abreu}, {Moll{\'a}}, {del Olmo}, {Meidt}, {Papaderos},
  {Puschnig}, {Quirrenbach}, {Roth}, {S{\'a}nchez-Bl{\'a}zquez}, {Spekkens},
  {Singh}, {Stanishev}, {Trager}, {Vilchez}, {Wild}, {Wisotzki}, {Zibetti}, \&
  {Ziegler}}]{Hus13}
{Husemann}, B., {Jahnke}, K., {S{\'a}nchez}, S.~F., {et~al.} 2013, \aap, 549,
  A87

\bibitem[{{Jarrett} {et~al.}(2017){Jarrett}, {Cluver}, {Magoulas}, {Bilicki},
  {Alpaslan}, {Bland-Hawthorn}, {Brough}, {Brown}, {Croom}, {Driver},
  {Holwerda}, {Hopkins}, {Loveday}, {Norberg}, {Peacock}, {Popescu}, {Sadler},
  {Taylor}, {Tuffs}, \& {Wang}}]{Jar17}
{Jarrett}, T.~H., {Cluver}, M.~E., {Magoulas}, C., {et~al.} 2017, \apj, 836,
  182

\bibitem[{{Jedrzejewski}(1987)}]{Jed87}
{Jedrzejewski}, R.~I. 1987, \mnras, 226, 747

\bibitem[{{Jogee}(2006)}]{Jog06}
{Jogee}, S. 2006, in Lecture Notes in Physics, Berlin Springer Verlag, Vol.
  693, Physics of Active Galactic Nuclei at all Scales, ed. D.~{Alloin}, 143

\bibitem[{{Jogee} {et~al.}(2005){Jogee}, {Scoville}, \& {Kenney}}]{Jog05}
{Jogee}, S., {Scoville}, N., \& {Kenney}, J.~D.~P. 2005, \apj, 630, 837

\bibitem[{{Kelz} {et~al.}(2006){Kelz}, {Verheijen}, {Roth}, {Bauer}, {Becker},
  {Paschke}, {Popow}, {S{\'a}nchez}, \& {Laux}}]{Kel06}
{Kelz}, A., {Verheijen}, M.~A.~W., {Roth}, M.~M., {et~al.} 2006, \pasp, 118,
  129

\bibitem[{{Kim} \& {Fabbiano}(2004)}]{kim04}
{Kim}, D.-W., \& {Fabbiano}, G. 2004, \apj, 613, 933

\bibitem[{{Knapen} {et~al.}(2000){Knapen}, {Shlosman}, \& {Peletier}}]{Kna00}
{Knapen}, J.~H., {Shlosman}, I., \& {Peletier}, R.~F. 2000, \apj, 529, 93

\bibitem[{{Kormendy}(1979)}]{Kor79}
{Kormendy}, J. 1979, \apj, 227, 714

\bibitem[{{Kormendy} \& {Kennicutt}(2004)}]{Kor04}
{Kormendy}, J., \& {Kennicutt}, Jr., R.~C. 2004, \araa, 42, 603

\bibitem[{{Larson} {et~al.}(1980){Larson}, {Tinsley}, \& {Caldwell}}]{Lar80}
{Larson}, R.~B., {Tinsley}, B.~M., \& {Caldwell}, C.~N. 1980, \apj, 237, 692

\bibitem[{{Lee} {et~al.}(2008){Lee}, {Lee}, {Park}, \& {Choi}}]{Lee08}
{Lee}, J.~H., {Lee}, M.~G., {Park}, C., \& {Choi}, Y.-Y. 2008, \mnras, 389,
  1791

\bibitem[{{Mahajan} \& {Raychaudhury}(2009)}]{Mah09}
{Mahajan}, S., \& {Raychaudhury}, S. 2009, \mnras, 400, 687

\bibitem[{{Marinova} \& {Jogee}(2007)}]{Mar07}
{Marinova}, I., \& {Jogee}, S. 2007, \apj, 659, 1176

\bibitem[{{Masters} {et~al.}(2010){Masters}, {Mosleh}, {Romer}, {Nichol},
  {Bamford}, {Schawinski}, {Lintott}, {Andreescu}, {Campbell}, {Crowcroft},
  {Doyle}, {Edmondson}, {Murray}, {Raddick}, {Slosar}, {Szalay}, \&
  {Vandenberg}}]{Mas10}
{Masters}, K.~L., {Mosleh}, M., {Romer}, A.~K., {et~al.} 2010, \mnras, 405, 783

\bibitem[{{Masters} {et~al.}(2012){Masters}, {Nichol}, {Haynes}, {Keel},
  {Lintott}, {Simmons}, {Skibba}, {Bamford}, {Giovanelli}, \&
  {Schawinski}}]{Mas12}
{Masters}, K.~L., {Nichol}, R.~C., {Haynes}, M.~P., {et~al.} 2012, \mnras, 424,
  2180

\bibitem[{{Men{\'e}ndez-Delmestre} {et~al.}(2007){Men{\'e}ndez-Delmestre},
  {Sheth}, {Schinnerer}, {Jarrett}, \& {Scoville}}]{Men07}
{Men{\'e}ndez-Delmestre}, K., {Sheth}, K., {Schinnerer}, E., {Jarrett}, T.~H.,
  \& {Scoville}, N.~Z. 2007, \apj, 657, 790

\bibitem[{{Mignoli} {et~al.}(2009){Mignoli}, {Zamorani}, {Scodeggio},
  {Cimatti}, {Halliday}, {Lilly}, {Pozzetti}, {Vergani}, {Carollo}, {Contini},
  {Le F{\'e}vre}, {Mainieri}, {Renzini}, {Bardelli}, {Bolzonella}, {Bongiorno},
  {Caputi}, {Coppa}, {Cucciati}, {de La Torre}, {de Ravel}, {Franzetti},
  {Garilli}, {Iovino}, {Kampczyk}, {Kneib}, {Knobel}, {Kova{\v c}},
  {Lamareille}, {Le Borgne}, {Le Brun}, {Maier}, {Pell{\`o}}, {Peng}, {Perez
  Montero}, {Ricciardelli}, {Scarlata}, {Silverman}, {Tanaka}, {Tasca},
  {Tresse}, {Zucca}, {Abbas}, {Bottini}, {Capak}, {Cappi}, {Cassata}, {Fumana},
  {Guzzo}, {Leauthaud}, {Maccagni}, {Marinoni}, {McCracken}, {Memeo}, {Meneux},
  {Oesch}, {Porciani}, {Scaramella}, \& {Scoville}}]{Mig09}
{Mignoli}, M., {Zamorani}, G., {Scodeggio}, M., {et~al.} 2009, \aap, 493, 39

\bibitem[{{Moore} {et~al.}(1999){Moore}, {Lake}, {Quinn}, \& {Stadel}}]{Moo99}
{Moore}, B., {Lake}, G., {Quinn}, T., \& {Stadel}, J. 1999, \mnras, 304, 465

\bibitem[{{Moran} {et~al.}(2006){Moran}, {Ellis}, {Treu}, {Salim}, {Rich},
  {Smith}, \& {Kneib}}]{Mor06}
{Moran}, S.~M., {Ellis}, R.~S., {Treu}, T., {et~al.} 2006, \apjl, 641, L97

\bibitem[{{Morandi} {et~al.}(2017){Morandi}, {Sun}, {Mulchaey}, {Nagai}, \&
  {Bonamente}}]{Mor17}
{Morandi}, A., {Sun}, M., {Mulchaey}, J., {Nagai}, D., \& {Bonamente}, M. 2017,
  \mnras, 469, 2423

\bibitem[{{O'Sullivan} {et~al.}(2017){O'Sullivan}, {Ponman}, {Kolokythas},
  {Raychaudhury}, {Babul}, {Vrtilek}, {David}, {Giacintucci}, {Gitti}, \&
  {Haines}}]{Osu17}
{O'Sullivan}, E., {Ponman}, T.~J., {Kolokythas}, K., {et~al.} 2017, \mnras,
  472, 1482

\bibitem[{{Parkash} {et~al.}(2018){Parkash}, {Brown}, {Jarrett}, \&
  {Bonne}}]{Par18}
{Parkash}, V., {Brown}, M.~J.~I., {Jarrett}, T.~H., \& {Bonne}, N.~J. 2018,
  ArXiv e-prints, arXiv:1807.06246

\bibitem[{{Poggianti} {et~al.}(1999){Poggianti}, {Smail}, {Dressler}, {Couch},
  {Barger}, {Butcher}, {Ellis}, \& {Oemler}}]{Pog99}
{Poggianti}, B.~M., {Smail}, I., {Dressler}, A., {et~al.} 1999, \apj, 518, 576

\bibitem[{{Puzia} {et~al.}(2005){Puzia}, {Kissler-Patig}, {Thomas}, {Maraston},
  {Saglia}, {Bender}, {Goudfrooij}, \& {Hempel}}]{Puz05}
{Puzia}, T.~H., {Kissler-Patig}, M., {Thomas}, D., {et~al.} 2005, \aap, 439,
  997

\bibitem[{{Rasmussen} {et~al.}(2012){Rasmussen}, {Bai}, {Mulchaey}, {van
  Gorkom}, {Jeltema}, {Zabludoff}, {Wilcots}, {Martini}, {Lee}, \&
  {Roberts}}]{Ras12}
{Rasmussen}, J., {Bai}, X.-N., {Mulchaey}, J.~S., {et~al.} 2012, \apj, 747, 31

\bibitem[{{Reese} {et~al.}(2007){Reese}, {Williams}, {Sellwood}, {Barnes}, \&
  {Powell}}]{Ree07}
{Reese}, A.~S., {Williams}, T.~B., {Sellwood}, J.~A., {Barnes}, E.~I., \&
  {Powell}, B.~A. 2007, \aj, 133, 2846

\bibitem[{{Rosales-Ortega}(2012)}]{Ros12}
{Rosales-Ortega}, F.~F. 2012, ArXiv e-prints, arXiv:1211.0277

\bibitem[{{Roth} {et~al.}(2005){Roth}, {Kelz}, {Fechner}, {Hahn}, {Bauer},
  {Becker}, {B{\"o}hm}, {Christensen}, {Dionies}, {Paschke}, {Popow}, {Wolter},
  {Schmoll}, {Laux}, \& {Altmann}}]{Rot05}
{Roth}, M.~M., {Kelz}, A., {Fechner}, T., {et~al.} 2005, \pasp, 117, 620

\bibitem[{{S{\'a}nchez} {et~al.}(2012){S{\'a}nchez}, {Kennicutt}, {Gil de Paz},
  {van de Ven}, {V{\'{\i}}lchez}, {Wisotzki}, {Walcher}, {Mast}, {Aguerri},
  {Albiol-P{\'e}rez}, {Alonso-Herrero}, {Alves}, {Bakos}, {Bart{\'a}kov{\'a}},
  {Bland-Hawthorn}, {Boselli}, {Bomans}, {Castillo-Morales}, {Cortijo-Ferrero},
  {de Lorenzo-C{\'a}ceres}, {Del Olmo}, {Dettmar}, {D{\'{\i}}az}, {Ellis},
  {Falc{\'o}n-Barroso}, {Flores}, {Gallazzi}, {Garc{\'{\i}}a-Lorenzo},
  {Gonz{\'a}lez Delgado}, {Gruel}, {Haines}, {Hao}, {Husemann},
  {Igl{\'e}sias-P{\'a}ramo}, {Jahnke}, {Johnson}, {Jungwiert}, {Kalinova},
  {Kehrig}, {Kupko}, {L{\'o}pez-S{\'a}nchez}, {Lyubenova}, {Marino},
  {M{\'a}rmol-Queralt{\'o}}, {M{\'a}rquez}, {Masegosa}, {Meidt},
  {Mendez-Abreu}, {Monreal-Ibero}, {Montijo}, {Mour{\~a}o}, {Palacios-Navarro},
  {Papaderos}, {Pasquali}, {Peletier}, {P{\'e}rez}, {P{\'e}rez}, {Quirrenbach},
  {Rela{\~n}o}, {Rosales-Ortega}, {Roth}, {Ruiz-Lara},
  {S{\'a}nchez-Bl{\'a}zquez}, {Sengupta}, {Singh}, {Stanishev}, {Trager},
  {Vazdekis}, {Viironen}, {Wild}, {Zibetti}, \& {Ziegler}}]{San12}
{S{\'a}nchez}, S.~F., {Kennicutt}, R.~C., {Gil de Paz}, A., {et~al.} 2012,
  \aap, 538, A8

\bibitem[{{S{\'a}nchez} {et~al.}(2016){S{\'a}nchez}, {Garc{\'{\i}}a-Benito},
  {Zibetti}, {Walcher}, {Husemann}, {Mendoza}, {Galbany}, {Falc{\'o}n-Barroso},
  {Mast}, {Aceituno}, {Aguerri}, {Alves}, {Amorim}, {Ascasibar},
  {Barrado-Navascues}, {Barrera-Ballesteros}, {Bekerait{\`e}},
  {Bland-Hawthorn}, {Cano D{\'{\i}}az}, {Cid Fernandes}, {Cavichia}, {Cortijo},
  {Dannerbauer}, {Demleitner}, {D{\'{\i}}az}, {Dettmar}, {de
  Lorenzo-C{\'a}ceres}, {del Olmo}, {Galazzi}, {Garc{\'{\i}}a-Lorenzo}, {Gil de
  Paz}, {Gonz{\'a}lez Delgado}, {Holmes}, {Igl{\'e}sias-P{\'a}ramo}, {Kehrig},
  {Kelz}, {Kennicutt}, {Kleemann}, {Lacerda}, {L{\'o}pez Fern{\'a}ndez},
  {L{\'o}pez S{\'a}nchez}, {Lyubenova}, {Marino}, {M{\'a}rquez},
  {Mendez-Abreu}, {Moll{\'a}}, {Monreal-Ibero}, {Ortega Minakata},
  {Torres-Papaqui}, {P{\'e}rez}, {Rosales-Ortega}, {Roth},
  {S{\'a}nchez-Bl{\'a}zquez}, {Schilling}, {Spekkens}, {Vale Asari}, {van den
  Bosch}, {van de Ven}, {Vilchez}, {Wild}, {Wisotzki}, {Y{\i}ld{\i}r{\i}m}, \&
  {Ziegler}}]{San16}
{S{\'a}nchez}, S.~F., {Garc{\'{\i}}a-Benito}, R., {Zibetti}, S., {et~al.} 2016,
  \aap, 594, A36

\bibitem[{{Sato} {et~al.}(2009){Sato}, {Matsushita}, {Ishisaki}, {Yamasaki},
  {Ishida}, \& {Ohashi}}]{Sat09}
{Sato}, K., {Matsushita}, K., {Ishisaki}, Y., {et~al.} 2009, \pasj, 61, S353

\bibitem[{{Sheth} {et~al.}(2005){Sheth}, {Vogel}, {Regan}, {Thornley}, \&
  {Teuben}}]{She05}
{Sheth}, K., {Vogel}, S.~N., {Regan}, M.~W., {Thornley}, M.~D., \& {Teuben},
  P.~J. 2005, \apj, 632, 217

\bibitem[{{Shlosman} {et~al.}(1990){Shlosman}, {Begelman}, \& {Frank}}]{Shl90}
{Shlosman}, I., {Begelman}, M.~C., \& {Frank}, J. 1990, \nat, 345, 679

\bibitem[{{Shlosman} {et~al.}(1989){Shlosman}, {Frank}, \& {Begelman}}]{Shl89}
{Shlosman}, I., {Frank}, J., \& {Begelman}, M.~C. 1989, \nat, 338, 45

\bibitem[{{Shlosman} {et~al.}(2000){Shlosman}, {Peletier}, \& {Knapen}}]{Shl00}
{Shlosman}, I., {Peletier}, R.~F., \& {Knapen}, J.~H. 2000, \apjl, 535, L83

\bibitem[{{Staveley-Smith} {et~al.}(2000){Staveley-Smith}, {Koribalski},
  {Stewart}, {Putman}, {Kilborn}, \& {Webster}}]{Sta00}
{Staveley-Smith}, L., {Koribalski}, B.~S., {Stewart}, I., {et~al.} 2000, in
  Astronomical Society of the Pacific Conference Series, Vol. 217, Imaging at
  Radio through Submillimeter Wavelengths, ed. J.~G. {Mangum} \& S.~J.~E.
  {Radford}, 50

\bibitem[{{Tabor} {et~al.}(2017){Tabor}, {Merrifield}, {Arag{\'o}n-Salamanca},
  {Cappellari}, {Bamford}, \& {Johnston}}]{Tab17}
{Tabor}, M., {Merrifield}, M., {Arag{\'o}n-Salamanca}, A., {et~al.} 2017,
  \mnras, 466, 2024

\bibitem[{{Thomas} {et~al.}(2003){Thomas}, {Maraston}, \& {Bender}}]{Tho03}
{Thomas}, D., {Maraston}, C., \& {Bender}, R. 2003, \mnras, 343, 279

\bibitem[{{Thomas} {et~al.}(2011){Thomas}, {Maraston}, \& {Johansson}}]{Tho11}
{Thomas}, D., {Maraston}, C., \& {Johansson}, J. 2011, \mnras, 412, 2183

\bibitem[{{Toomre} \& {Toomre}(1972)}]{Too72}
{Toomre}, A., \& {Toomre}, J. 1972, \apj, 178, 623

\bibitem[{{Trager} {et~al.}(1998){Trager}, {Worthey}, {Faber}, {Burstein}, \&
  {Gonz{\'a}lez}}]{Tra98}
{Trager}, S.~C., {Worthey}, G., {Faber}, S.~M., {Burstein}, D., \&
  {Gonz{\'a}lez}, J.~J. 1998, \apjs, 116, 1

\bibitem[{{Tully}(2015)}]{Tul15}
{Tully}, R.~B. 2015, \aj, 149, 171

\bibitem[{{van den Bergh}(1976)}]{van76}
{van den Bergh}, S. 1976, \apj, 206, 883

\bibitem[{{Vazdekis} {et~al.}(2010){Vazdekis}, {S{\'a}nchez-Bl{\'a}zquez},
  {Falc{\'o}n-Barroso}, {Cenarro}, {Beasley}, {Cardiel}, {Gorgas}, \&
  {Peletier}}]{Vaz10}{Vazdekis}, A., {S{\'a}nchez-Bl{\'a}zquez}, P., {Falc{\'o}n-Barroso}, J.,
  {et~al.} 2010, \mnras, 404, 1639

\bibitem[{{Walcher} {et~al.}(2014){Walcher}, {Wisotzki}, {Bekerait{\'e}},
  {Husemann}, {Iglesias-P{\'a}ramo}, {Backsmann}, {Barrera Ballesteros},
  {Catal{\'a}n-Torrecilla}, {Cortijo}, {del Olmo}, {Garcia Lorenzo},
  {Falc{\'o}n-Barroso}, {Jilkova}, {Kalinova}, {Mast}, {Marino},
  {M{\'e}ndez-Abreu}, {Pasquali}, {S{\'a}nchez}, {Trager}, {Zibetti},
  {Aguerri}, {Alves}, {Bland-Hawthorn}, {Boselli}, {Castillo Morales}, {Cid
  Fernandes}, {Flores}, {Galbany}, {Gallazzi}, {Garc{\'{\i}}a-Benito}, {Gil de
  Paz}, {Gonz{\'a}lez-Delgado}, {Jahnke}, {Jungwiert}, {Kehrig}, {Lyubenova},
  {M{\'a}rquez Perez}, {Masegosa}, {Monreal Ibero}, {P{\'e}rez}, {Quirrenbach},
  {Rosales-Ortega}, {Roth}, {Sanchez-Blazquez}, {Spekkens}, {Tundo}, {van de
  Ven}, {Verheijen}, {Vilchez}, \& {Ziegler}}]{Wal14}
{Walcher}, C.~J., {Wisotzki}, L., {Bekerait{\'e}}, S., {et~al.} 2014, \aap,
  569, A1

\bibitem[{{Walker} {et~al.}(1996){Walker}, {Mihos}, \& {Hernquist}}]{Wal96}{Walker}, I.~R., {Mihos}, J.~C., \& {Hernquist}, L. 1996, \apj, 460, 121

\bibitem[{{Weinberg}(1985)}]{Wei85}{Weinberg}, M.~D. 1985, \mnras, 213, 451

\bibitem[{{Wolf} {et~al.}(2009){Wolf}, {Arag{\'o}n-Salamanca}, {Balogh},{Barden}, {Bell}, {Gray}, {Peng}, {Bacon}, {Barazza}, {B{\"o}hm}, {Caldwell}, {Gallazzi}, {H{\"a}u{\ss}ler}, {Heymans}, {Jahnke}, {Jogee}, {van Kampen},  {Lane}, {McIntosh}, {Meisenheimer}, {Papovich}, {S{\'a}nchez}, {Taylor},
  {Wisotzki}, \& {Zheng}}]{Wol09}{Wolf}, C., {Arag{\'o}n-Salamanca}, A., {Balogh}, M., {et~al.} 2009, \mnras, 393, 1302

\bibitem[{{Worthey} \& {Ottaviani}(1997)}]{Wor97}{Worthey}, G., \& {Ottaviani}, D.~L. 1997, \apjs, 111, 377

\bibitem[{{Yamauchi} \& {Goto}(2004)}]{Yam04}{Yamauchi}, C., \& {Goto}, T. 2004, \mnras, 352, 815


\end{thebibliography}
\end{document}